\documentclass[aps,prl,preprint,groupedaddress,showpacs]{revtex4-2}
\usepackage{graphicx,amsmath}
\begin{document}

\title{Quasipotentials in the Nonequilibrium Stationary States or a method to get explicit solutions of Hamilton-Jacobi equations}
\author{P.L. Garrido}
\email[]{garrido@onsager.ugr.es}
\affiliation{Instituto Carlos I de F{\'\i}sica Te{\'o}rica y Computacional. Universidad de Granada. E-18071 Granada. Spain }
\begin{abstract}
We assume that a system at a mesoscopic scale is described by a field $\phi(x,t)$ that evolves by a Langevin equation with a white noise whose intensity is controlled by a parameter $1/\sqrt{\Omega}$. The system stationary state distribution  in the small noise limit ($\Omega\rightarrow\infty$) is of the form $P_{st}[\phi]\simeq\exp(-\Omega V_0[\phi])$ where $V_0[\phi]$ is called the {\it quasipotential}. $V_0$ is the unknown of a Hamilton-Jacobi equation. Therefore, $V_0$ can be written as an action computed along a path that is the solution from Hamilton's equation that typically cannot be solved explicitly. This paper presents a theoretical scheme that builds a suitable canonical transformation that permits us to do such integration by deforming the original path into a straight line and including some weights along with it. We get the functional form of such weights through conditions on the existence and structure of the canonical transformation. We apply the scheme to get the quasipotential algebraically for several one-dimensional nonequilibrium models as the diffusive and reaction-diffusion systems.
\end{abstract}
\date{\today}
\maketitle

\section{I. Introduction}

Thermodynamics shows that many macroscopic properties of systems at equilibrium states are related to each other through a Thermodynamic Potential. Once we know, for instance, the Entropy for a one-component system as a function of energy and mass density, we can deduce many other observables: specific heat, compressibility, Pressure, or Temperature. However, Thermodynamics does not give us the explicit form of a system's Thermodynamic Potential. The Equilibrium Statistical Mechanics solves this problem by introducing the Gibbs invariant measure for the microscopic degrees of freedom. In our opinion, the elegant part of this connection between the microscopic and macroscopic descriptions is that the Gibbs measure depends on the object that defines the full microscopic system dynamics: the hamiltonian. Therefore, all the system's dynamical microscopic details are summed up and contained in the Thermodynamic Potentials. We see that systems at equilibrium have a complete set of theories that allow us to address much interesting macroscopic behavior, for instance, phase transitions. 

However, systems at equilibrium are not the most common states in Nature. Typically the systems at stationary states contain currents of any type, energy, mass,...  that appear due to unbalanced boundary conditions and/or the effect of external agents that induce some driving. From the microscopic point of view, few things, but very relevant ones, have changed compared with the equilibrium case: we have typically a system of interacting particles whose dynamics is still Hamiltonian except that we now include on it the dynamical effects from the boundaries and/or the external agents. These apparently small changes break down the theories that apply to systems at equilibrium. First, we do not know how to build a complete macroscopic theory similar to Thermodynamics. Nevertheless, there have been many efforts to justify the existence of some intermediate or mesoscopic descriptions as the Boltzmann equation or macroscopic ones as the Navier-Stokes equations for fluids \cite{Spohn}. And second,  the natural invariant measure defined on the phase space is of no practical use. For instance, we can use the  SRB measure at the non-equilibrium attractor when the system is ``very'' chaotic \cite{Young}. It is expected that the volume of the nonequilibrium attractor is zero due to dissipation. Still, it could be assumed dense in phase space when the degrees of freedom tend to infinity. However, the attractor's topological structure typically depends strongly on the overall system dynamic trajectories, and therefore it is unknown \cite{Gallavotti3}. We should compare this complex structure with, for example, the microcanonical measure at equilibrium where it is constant on the ``a priori'' well-known attractor that is the equal energy manifold $H(x,p)=E$. 

Some of these problems may be circumvented by studying systems with markovian dynamics. The attractors are compact sets that depend on the physical constraints of the variables. Therefore many of the complexities associated with the attractor topology go away compared with nonequilibrium particle systems.  There have been many efforts to elucidate general properties of such non-equilibrium systems on lattice models: voter model, contact process, exclusion process,... \cite{Liggett}.  For many years the stationary measure typically could only be obtained in few simple cases as the zero-dimensional stochastic models, systems with local detailed balance condition \cite{Gardiner} or in the thermodynamic limit of the KMP model for heat conduction \cite{KMP}. A breakthrough took place by the rigorous derivation of the stationary probability for the one-dimensional boundary driven Symmetric Simple Exclusion model (SSE) by Derrida et al. (2001) \cite{Derrida}.  In SSE, each site can be empty or with one particle. The dynamics is very simple: a randomly chosen particle may hop to an empty neighbor's site with some given probability. However, at the two boundaries, there are exit and incoming probability rates that may be different. Therefore, it may be created a net current of particles through the system. When the number of lattice sites, $N$, tend to infinity and the density at the boundaries is fixed and given by $\phi(0)=\phi_0$ and $\phi(1)=\phi_1$, $\phi_0>\phi_1$, Derrida et al. deduce that the probability to find a density profile $\eta=\{\eta(x),\forall x\in[0,1]\}$, where $x\in [0,1]$ and $\eta(x)\in[0,1]$, is given by a large deviation functional:
 \begin{equation}
 P[\eta]\simeq\exp\left[-N(V_0[\eta])\right]\quad (N\rightarrow\infty)
 \end{equation}
where $V_0[\eta]$ is called {\it quasipotential} and it is given by:
\begin{equation}
V_0[\eta]=V_0[\phi^*]+G[\eta,\tilde\eta] \label{ss1}
\end{equation}
with 
\begin{equation}
G[\eta,\tilde\eta]=\int_0^1dx\biggl[\eta(x)\log\left[\frac{\eta(x)}{\tilde\eta(x)}\right] +(1-\eta(x))\log\left[\frac{1-\eta(x)}{1-\tilde\eta(x)}\right]
+\log\left[\frac{\tilde\eta'(x)}{\phi_1-\phi_0}\right]
 \biggr]\label{ss2}
\end{equation}
where $\tilde\eta(x)$ is an auxiliary function that is solution of the second order differential equation
\begin{equation}
\eta(x)=\tilde\eta(x)+\tilde\eta(x)(1-\tilde\eta(x))\frac{\tilde\eta''(x)}{\tilde\eta'(x)^2}\label{de}
\end{equation}
with boundary conditions: $\tilde\eta(0,1)=\phi_{0,1}$ and the stationary profile is given by $\phi^*(x)=\phi_0+x(\phi_1-\phi_0)$.
They also mention that all the $\tilde\eta(x)$ functions such that
\begin{equation}
\frac{\delta G[\eta,\tilde\eta]}{\delta\tilde\eta(x)}=0\label{ss3}
\end{equation}
are the ones that solve the differential equation (\ref{de}) which is a bite intriguing.  The reader can find at Ref.\cite{Derrida} many interesting properties and insights of this explicit quasipotential as, for instance, the existence, uniqueness, and monotonicity of the solutions $\tilde\eta(x)$ from the differential equation (\ref{de}).

Let us remind that the quasipotential for systems at equilibrium is directly related to the free energy functional that is a {\it Thermodynamic Potential}. That makes $V_0$ to be a fascinating object to analyze when looking for a non-equilibrium thermodynamic theory (if possible). For instance, let us focus just on the mathematical structure of Derrida's result. Please observe that the quasipotential seems to be a local functional as it happens at equilibrium. However, the auxiliary field that is the solution of the second-order differential equation depends on the given $\eta(x)$ in a non-local way and on the boundary conditions. In our opinion, the elegant and inspiring part of this solution is how the non-local behavior that is typical in many non-equilibrium stationary states is mathematically codified. 

Later, using similar techniques, Enaud and Derrida \cite{Enaud} obtained the quasipotential for the boundary-driven asymmetric simple exclusion process (ASEP) for the driving field aligned with the density gradient. They obtained a quasipotential's mathematical form similar to the SSEP case above showed: a local functional depending on an auxiliary field and a second-order differential equation for it. The strategy of studying lattice models with markovian dynamics was successful, and it gave us an important reference on the structure of the quasipotentials. However, up to our knowledge, these results based on Derrida's matrix technique for one-dimensional models with exclusion process are the unique ones where the quasipotential have been exactly derived. 

A simplified formal path towards the quasipotentials study was already on the stake by using Fokker-Planck descriptions of non-equilibrium situations \cite{Gardiner}. In the context of non-equilibrium many-body systems, it is assumed that there are a set of macroscopic fields that evolve following a known deterministic dynamics and a weak stochastic term (typically white noise) reminiscence of the microscopic fluctuations. Let us mention the pioneering work of Graham et al. \cite{Graham0} where it is studied the general properties of the quasipotential for systems with finite degrees of freedom. For instance, they introduced the possibility of a  Lagrangian transition that would imply the non-differentiability of the quasipotential in some regions of configurational space. In fact, this property seems to be natural for many systems at non-equilibrium stationary states. Let us also point out that they also developed a gradient expansion of the quasipotential for the supercritical complex Ginzburg-Landau equation \cite{Descalzi}. These works have been applied with great success in several models and fields. Let us remark here just its use in the study of biological systems where the quasipotentials give a complete description of, for instance, the most probable path that a complex network of chemical reactions follows  to go from a local minimum to another one \cite{Fang, Wang}. Let us finally mention the seminal works from Donkser and Varadhan \cite{Donkser} that developed the mathematical Theory of Large Deviations (TLD) for Markov processes.  TLD gives mathematical support on the existence and properties of the rate function that, in our context, is the quasipotential (see the interesting review by Touchette about TDL in Statistical Mechanics \cite{Hugo}).

Bertini and coworkers introduced further improvement by formulating the Macroscopic Fluctuation Theory (MFT) \cite{Bertini0}. MFT was formulated based on many previous rigorous results connecting microscopic stochastic lattice models with their corresponding macroscopic dynamical equations. Large deviation formulas were also obtained, and thus, the mesoscopic description of such systems. They compiled all this information to define a theory that is a generalization of the already known fluctuating hydrodynamics \cite{Fox}. That is, systems described by hydrodynamic continuum fields that evolve following a Langevin-like equation. First, they computed the quasipotential for the zero-range model, and they found that it is local. Moreover, they applied MFT to the continuum mesoscopic version of the SSE model. They found that the quasipotential obtained by Derrida et al. (Eqs. (\ref{ss1}) and \ref{ss2}) was also the solution of the Hamilton-Jacobi equation that defined the quasipotential in MFT.  Bertini et al. \cite{Bertini1} obtained the quasipotential for the mesoscopic version of the Kipnis, Marchioro, Presutti model for heat conduction (KMP)\cite{KMP}. In fact, they proposed a functional $G[\eta,\tilde\eta]$ inspired by eq.(\ref{ss2}), and they showed that it was the solution of the corresponding Hamilton-Jacobi equation from MFT. They also found that the quasipotential for the boundary driven ASEP from Enaud and Derrida \cite{Enaud} was also the solution for the MFT \cite{Bertini2} and expanded such result when the drift due to the external field is strong enough, and it points against to that due to the density gradient \cite{Bertini3}. They explicitly found a Lagrangian Transition, that is, the quasipotential has non-differential behavior in this case. All these results showed that MFT had solid theoretical grounds to describe non-equilibrium systems at the mesoscopic level correctly. Let us mention that the quasipotentials study is just a part of the set of properties of non-equilibrium systems that MFT describes self-consistently.  A fascinating review of many aspects that MFT  sheds some light on can be found in ref.\cite{Bertini}.

We have seen that the exact results from Derrida et al. using their matrix method and the inspired works from Bertini et al. by defining MFT have open the way for a deep understanding of the quasipotential's mathematical structure. However, to go beyond this point, it is needed new insights that permit us to study more systematically other systems or/and dimensions. This paper focuses on looking for a general algebraic method to obtain the quasipotential from MFT. 

As we will see, in the MFT context, the quasipotential is solution of a  Hamilton-Jacobi equation of the form:
\begin{equation}
H[\eta,\frac{\delta V_0[\eta]}{\delta\eta}]\equiv\int_\Lambda dx \,\bar h\left[\eta(x),\frac{\delta V_0[\eta]}{\delta\eta(x)}\right] =0
\end{equation}
where $\bar h$ is a local functional on the arguments. Formally, this equation is solved by using the method of characteristics \cite{Gall}. That is, $H[\eta,\pi]$ is assumed to be a hamiltonian that defines a dynamical system where $\pi(x)$ is the conjugate field to $\phi(x)$. The quasipotential $V_0[\eta]$ is then given by:
\begin{equation}
V_0[\eta]=V_0[\phi^*]+
\int_{-\infty}^0 dt\int_\Lambda dx\, \pi(x,t)\partial_t\phi(x,t)
\end{equation}
where the fields $(\phi(x,t),\pi(x,t))$ are solution of the Hamilton's equations associated to the hamiltonian $H[\phi,\pi]$ where $(\phi(x,-\infty),\pi(x,-\infty))=(\phi^*(x),0)$ and $(\phi(x,0),\pi(x,0))=(\eta(x),\pi(x))$.

Except for trivial cases, it is unknown how to solve the hamilton equations to get the trajectory $(\phi(t),\pi(t))$. Therefore we are unable to compute the time integral to get $V_0$. At this point,  the need to solve the Hamilton equations induced us to ask the following question: Is it possible to define a canonical transformation $(\phi,\pi)\rightarrow (\tilde\phi,\tilde\pi)$ such that, in the new variables, we can make the time-integral to get $V_0$? 

In this paper, we explore this idea by using a type 1
canonical transformation defined by the generator of the transformation $L[\phi,\tilde\phi]$. It is impossible to write down a $L$ generator such that in the new variables, we could do the time integrals explicitly. Therefore, we first assume the existence of a map $\phi(x)=\phi[\tilde\phi;x]$ between the hamiltonian paths at each $t$. We show that under this assumption, the quasipotential can be obtained by a parametric integral that connects, by a straight line, the stationary state $\phi^*$ with $\eta$. However, the integral is now weighted by two unknown functionals. We assume explicit analytical forms for the unknown functionals. We determine them from a set of compatibility conditions to fulfill to be part of a well-defined canonical transformation.
As we will see, the interesting part of this method is that we do not need to solve any differential equation. This scheme is imperfect because we should also restrict the original dynamical model to some concrete forms for each trial functionals form. Moreover, not always the pair, model, and elected functionals have a solution, and therefore this is a digging-like method to find some gold nuggets. Nevertheless, we reproduce all the known quasipotentials with this method, and we discover some new ones.

We present all these results in the following manner.
In section II, we define the Langevin dynamics of the system and fast review how to get the quasipotential and some concepts that we will use. Moreover, we define the one-dimensional models we are study explicitly in the paper. In section III, we do the canonical transformation and see how the quasipotential's formal solution is affected. We introduce the necessity for the map between $\phi$ and $\tilde\phi$. We derive the quasipotential that appears to depend on two unknown functionals. We find the conditions they should follow to be part of a well-defined canonical transformation. Finally, as an example, we get the quasipotential for the zero-range model by using our method. Section IV generalizes the method by defining some general functional forms for the unknown functionals and expressing their compatibility conditions on operational form. Section V is devoted to obtaining quasipotentials using our method for the one-dimensional diffusive system. In section VI, we derive some quasipotentials for one-dimensional reaction-diffusion models. 

\section{II. The quasipotential for a Langevin description of mesoscopic systems and models studied}

We assume that our systems at a mesoscopic level of description are characterized by a unique scalar field $\phi(x,t)\in {\rm I\!R}$ where $x\in \Lambda \subset {\rm I\!R}^d$, $d$ is the spatial dimension and $t$ is the time. We have initially restricted ourselves to this case in this paper for the sake of simplicity. Still, one can straightforward generalize all the results below to systems described by vector fields. The system dynamics is given by a mesoscopic Langevin equation with a white noise. For instance, in the case of a reaction dynamics (RD) it is:
\begin{equation}
\partial_t\phi(x,t)=F[\phi(t);x]+h[\phi(t);x]\xi(x,t)\label{rd}
\end{equation}
where we use in this paper the notation $W[\phi,\psi,\ldots;x]$ to indicate a local functional that may depend on the fields:  $\phi(x)$, $\psi(x)$, their first derivatives with respect the argumens $x$: $\nabla_x\phi(x)$, $\nabla_x\psi(x)$,  higher derivatives and even they may have more complex structures as local integrals of the fields over domains around $x$. Moreover, a parametric dependence on the fields, for instance the time $t$, is written as $W[\phi(t);x]$ meaning that $W$ is a local functional that depend on $\phi(x,t)$, their derivatives on $x$ or any other $x$-functional dependence. Finally, $\xi(x,t)$ is an uncorrelated gaussian random field:
\begin{eqnarray}
\langle\xi(x,t)\rangle&=&0\nonumber\\
\langle\xi(x,t)\xi(x',t')\rangle&=&\frac{1}{\Omega}\delta(x-x')\delta(t-t')
\end{eqnarray}
and we follow the Ito's scheme.
The dynamics becomes deterministic when $\Omega\rightarrow\infty$:
\begin{equation}
\partial_t\phi_D(x,t)=F[\phi_D(t);x]\label{det_rd}
\end{equation}
We assume along this paper that the deterministic dynamics has a unique stationary state and that it is locally stable:
\begin{equation}
 F[\phi^*;x]=0\quad, \quad \phi^*(x)=\lim_{t\rightarrow\infty}\phi_D(x,t)\label{ss}
 \end{equation}
 for almost any initial state $\phi_D(x,0)=\phi_0(x)\in\Lambda$. Our system may have periodic boundary conditions ($\phi(x+L)=\phi(x)$ with $L$ being the vector defining the basic cell),  fix boundary conditions ($\phi(x,t)=f_0(x)$, $\forall x\in\partial\Lambda$) or a mixture of both. 

When the noise intensity is very small, the stationary probability distribution is of the form:
\begin{equation}
P_{st}[\eta]\simeq \exp\left[-\Omega V_0[\eta]\right] \quad (\Omega\rightarrow\infty)\label{v0}
\end{equation}
where $V_0[\eta]$ is the so-called {\it quasipotential}. It is welll known \cite{Graham0,Garrido0} that $V_0$ is solution of the {\it Hamilton-Jacobi equation}:
\begin{equation}
H\left[\phi,\frac{\delta V_0[\phi]}{\delta\phi}\right]\equiv 
\int_\Lambda dx\,\frac{\delta V_0[\phi]}{\delta\phi(x)}\left[F[\phi;x]+\frac{1}{2}\frac{\delta V_0[\phi]}{\delta\phi(x)}h[\phi;x]^2\right]
=0\label{hamRD}
\end{equation}
with boundary conditions:
\begin{equation}
\frac{\delta V_0[\phi]}{\delta\phi(x)}\biggr\vert_{ x\in\partial\Lambda}=0
\end{equation}
The formal solution can be obtained from the expression:
\begin{equation}
V_0[\eta]=V_0[\phi^*]+
\int_{-\infty}^0 dt\int_\Lambda dx\, \pi(x,t)\partial_t\phi(x,t)\label{action2}
\end{equation}
where the fields $(\pi(x,t),\phi(x,t))$ are solution of the Hamilton's equations:
\begin{eqnarray}
\partial_t\phi(x,t)&=&\frac{\delta H[\phi(t),\pi(t)]}{\delta\pi(x,t)}\nonumber\\
\partial_t\pi(x,t)&=&-\frac{\delta H[\phi(t),\pi(t)]}{\delta\phi(x,t)}\label{Heq}
\end{eqnarray}
where the hamiltonian $H$ is defined by eq.(\ref{hamRD}).
Hamilton's equations should be solved with the system's spatial boundary conditions and with the time boundaries: $(\phi(x,-\infty),\pi(x,-\infty))=(\phi^*(x),0)$ and $(\phi(x,0),\pi(x,0))=(\eta(x),\pi(x))$ $\forall x\in\Lambda$. 

Let us point out some properties that we will use below:
\begin{itemize}
\item  $\phi^*(x)$ is the absolute minimum of the quasi potential: 
\begin{equation}
\frac{\delta V_0[\phi]}{\delta\phi(x)}\biggr\vert_{\phi(x)=\phi^*(x)}=0\quad\forall x\in\Lambda
\end{equation}
That is so because in the strict limit $\Omega\rightarrow\infty$ we should get the stationary deterministic solution (\ref{ss}). In other words:
\begin{equation}
P_{st}[\eta]=\prod_{x\in\Lambda}\delta(\eta(x)-\phi^*(x))
\end{equation}

\item $H[\phi^*,0]=0$ by construction and therefore, $H[\phi(t),\pi(t)]=0$.

\item $\pi(x,t)=\delta V_0[\eta]/\delta\eta(x)\vert_{\eta=\phi(t)}$  from eq.(\ref{action2}) .

\item Notice that for fix boundary conditions:  $(\phi(x,t),\pi(x,t))=(\phi(x),0)$ $\forall x\in\partial\Lambda,\, t$. Where we have included $\pi(x,t)\vert_{x\in\partial\Lambda}=0$. This condition reflects that the boundary is thought as an equilibrium thermal bath having the property $\partial V_B(\phi)/\partial\phi(x)=0$, with $V_B$ an equilibrium potential. That is, $\pi_B(x)\vert_{x\in\partial\Lambda}=0$ and, by continuity $\pi_B(x)=\pi(x)$ $\forall x\in\partial\Lambda$. This choice, of course, affects the nature of the fluctuations about the system's stationary state, but it has the advantage that it permits us to have systems at equilibrium. Then, just by changing the boundaries, we can create nonequilibrium stationary states. 

\item There can be more than one path solution of Hamilton's equations that go from $(\phi^*,0)$ up to  $(\eta,\pi_n)$ where $n$ could change for each of the paths. Then it is implicitly understood that one should take in eq. (\ref{action2}) the path that minimizes the value of $V_0$. 

\item For Diffusive Dynamics (DD) everything is equal except for the Hamiltonian (\ref{hamRD}) that  in this case is:
\begin{equation}
H[\phi,\pi]=\int_\Lambda dx\,\nabla\pi(x)\cdot\left[G[\phi;x]+\frac{1}{2}\chi[\phi;x]\nabla\pi(x)\right]\label{hamDD}
\end{equation}
where $G$ is the determinist part of the current and $\chi$ is related with the noise intensity.
\end{itemize}
The above definitions and properties are well known in the literature, and that's why we pass through them fast. We refer the readers to ref.\cite{Garrido0} for the details about how the above expressions are derived for systems with RD and DD  and several comments about the properties of the stationary state.

That is, the problem of finding $V_0$ is formally solved. However, it is almost impossible at the practical level to obtain the solutions from Hamilton's equations (\ref{Heq}). This paper is devoted to building a strategy to be able to make explicitly the time-integral in eq. (\ref{action2}).  From now on, we are going to restrict ourselves to one-dimensional systems. The application of these ideas to larger dimensions is left for future works. Below, we apply explicitly the method to get $V_0$ to the following one dimensional systems:

\begin{itemize} 

\item {\bf Diffusive Model:}

This one dimensional model is defined by a field $\phi(x,t)$ with $x\in[0,1]$ that evolves by  the Langevin equation:
\begin{equation}
\partial_t\phi(x,t)+\frac{d j[\phi(t);x]}{dx}\quad,\quad j[\phi(t);x]=G[\phi(t);x]+\sqrt{\chi(\phi(x,t))}\psi(x,t)\label{Dmodel1}
\end{equation}
where $\psi$ is a uncorrelated white noise and
\begin{equation}
G[\phi;x]=-D(\phi(x))\frac{d\phi}{dx}+\chi(\phi(x)) E \label{Dmodel2}
\end{equation} 
$D(\lambda)$ and $\chi(\lambda)$ are the diffusion and mobility functions respectively and $E$ is a constant driving field.
The hamiltonian that define the paths to build the quasipotential is given by eq.(\ref{hamDD})
(see Ref.\cite{Garrido0} and references therein):
\begin{equation}
H[\phi,\pi]=\int_\Lambda dx\, \frac{d\pi(x)}{dx}\left[-D(\phi(x))\frac{d\phi(x)}{dx}+\chi(\phi(x))\left(E+\frac{1}{2}\frac{d\pi(x)}{dx}\right)\right]\label{Honedif}
\end{equation}
$D$ and $\chi$ are designed in such a way that the system stationary state could be an equilibrium state with respect the potential:
\begin{equation}
V_{eq}[\phi]=V_{eq}[\phi^*]+\int_{0}^1dx\,\left[v_{eq}[\phi;x]-v_{eq}[\phi^*;x]\right]
\end{equation}
where
\begin{equation}
v_{eq}[\phi;x]=v(\phi(x))-2Ex\phi(x)
\end{equation}
We can think $E$ being a kind of gravitational force acting over a mass field $\phi(x)$. 
We can see that $V_{eq}[\phi]$ is the solution of the Hamilton-Jacobi equation
\begin{equation}
H[\phi,\frac{\delta V_{eq}[\phi]}{\delta\phi}]=0
\end{equation}
when
\begin{equation}
D(\lambda)=\frac{1}{2}v''(\lambda)\chi(\lambda)\label{er}
\end{equation}
that it is called {\it Einstein Relation}.
 The equilibrium state is achieved when applying the appropriate boundary conditions:
\begin{equation}
\frac{\delta V_{eq}[\phi]}{\delta\phi(x)}\biggr\vert_{x=0,1}=0 \quad\Rightarrow\quad v'(\phi_0)=0\quad ,\quad v'(\phi_1)=2E
\end{equation}
where $\phi(i)=\phi_i\quad i=0,1$.

Finally, the equilibrium configuration is obtained from the deterministic part of the Langevin equation by asking that the current $G$ equals to zero:
\begin{equation}
-D(\phi^*(x))\frac{d\phi^*(x)}{dx}+\chi(\phi^*(x))E=0
\end{equation}
The solution of this equation is
\begin{equation}
\int_{\phi(0)}^{\phi^*(x)}d\phi\,\frac{D(\phi)}{\chi(\phi)}=Ex
\end{equation}
and assuming that the Einstein relation holds, it can be written
\begin{equation}
v'(\phi^*(x))=2Ex
\end{equation}
where $\phi^*(i)=\phi_i\quad i=0,1$. Observe that the boundary conditions should be $\phi_{0,1}$ (for a given $E$) to be at an equilibrium state. When we choose any other different set of boundary conditions, the system develops a non-zero current, and the system is in a non-equilibrium stationary state with a quasipotential $V_0[\phi]\neq V_{eq}[\phi]$. The stationary state is then solution of
\begin{equation}
-D(\phi^*(x))\frac{d\phi^*(x)}{dx}+\chi(\phi^*(x))E=J
\end{equation}
where $J$ is the current that it is determined by the boundary conditions.
 We can also get non-equilibrium stationary states with periodic boundary conditions and a non-zero driving field $E$. In this case $\phi^*(x)=\phi^*$ and $J=\chi(\phi^*)E$. 

\item {\bf Reaction-Diffusive models:}

We study the one-dimensional reaction-diffusion model whose  Langevin equation is given by (\ref{rd}) with
\begin{equation}
F[\phi;x]=g(\phi)\phi''(x)+w(\phi(x))
\end{equation}
with the hamiltonian given by eq.(\ref{hamRD}).

Another interesting model we have studied is the Poissonian Reaction-Diffusion Dynamics. This mesoscopic model is deduced from a stochastic markovian lattice model in which there is a competition between conservative exchange dynamics and a spin-flip one (see ref.\cite{anna}). In the fast rate limit for the exchange dynamics and after some time and space rescaling, one  obtains the deterministic equation:
\begin{equation}
\frac{\partial\phi_D(x,t)}{\partial t}=\frac{\partial^2\phi_D(x,t)}{\partial x^2}+b(\phi_D(x,t))-d(\phi_D(x,t))
\end{equation}
where  $\phi_D(x)$ represents a normalized density: $0\leq\phi_D(x)\leq 1$ and $b$ and $d$ functions are directly related with the microscopic spin-flip dynamics. Moreover, the structure of the mesoscopic noise is represented by the hamiltonian:
\begin{eqnarray}
H[\phi,\pi]&=&\int_\Lambda dx\biggl[\pi(x)\phi''(x)+\pi'(x)\phi(x)(1-\phi(x))\nonumber\\
&-&b(\phi(x))(1-e^{\pi(x)})-d(\phi(x))(1-e^{-\pi(x)}) \biggr]
\end{eqnarray}
Observe that this hamiltonian is not quadratic in $\pi$ as it was in eq.(\ref{hamRD}). That is related to the Poissonian structure of the underlying noise.  We will assume periodic boundary conditions. In this case the stationary state is a constant solution of $b(\phi^*)=d(\phi^*)$ that it is assumed to be unique.
\end{itemize}

\section{III. A method to solve Hamilton-Jacobi equations}

We want to get explicit solutions for the Hamilton-Jacobi equation (\ref{hamRD}) by using the formal solution (\ref{action2}) as starting point. The main idea of our method is to find a canonical transformation under which we can explicitly do time integral in (\ref{action2}). We'll see that we can deform the integrating path to be, effectively, a straight line. We can do that under some (assumed mild) assumptions, such as the existence of a one-to-one transform between the original path and the new one coming from the canonical transformation.  However, the transformed action integral (\ref{action2}) is weighted by two functionals. One is the canonical transformation's functional derivative, and the other depends on the one-to-one transform. The key point of the method is to get both unknown functionals. We show that those functionals can be algebraically obtained by using the conditions about the existence of the canonical transformation and the use of the Hamiltonian equation of motion for the original and canonical transformed systems. Once we get such weights, we can compute straightforward the quasipotential. Let us develop the full strategy step by step.

\subsection{III.1. The canonical transformation}

 Let us build a general {\it Type 1} canonical transformation on a generic field hamiltonian $H(\phi,\pi)$ through the generator $L[\phi,\tilde\phi]$:
\begin{equation}
(\phi,\pi)\rightarrow (\tilde\phi,\tilde\pi)\quad :\quad \pi(x)=\frac{\delta L[\phi,\tilde\phi]}{\delta\phi(x)}\equiv A[\phi,\tilde\phi;x]\quad,\quad \tilde\pi(x)=-\frac{\delta L[\phi,\tilde\phi]}{\delta\tilde\phi(x)}\equiv B[\phi,\tilde\phi;x]\label{ct}
\end{equation}
These equations define a one-to-one relationship between the two sets of variables during the system's evolution under the hamiltonian $H$. The quasipotential (\ref{action2}) is written in the new variables:
\begin{equation}
V_0[\eta]=V_0[\phi^*]+\int_{-\infty}^0 d\tau\,\int_\Lambda dx\,\frac{\delta L[\phi,\tilde\phi]}{\delta\phi(x)}\biggr\vert_{\substack{\phi=\phi(\tau)\\ \tilde\phi=\tilde\phi(\tau)}}\partial_\tau\phi(x,\tau) 
\end{equation}
$L$ does not depend on $t$ explicitly, and therefore we can use the relation:
\begin{equation}
\partial_t L[\phi(t),\tilde\phi(t)]=\int_{\Lambda}dx\, \left[\frac{\delta L[\phi,\tilde\phi]}{\delta\phi(x)}\biggr\vert_{\substack{\phi=\phi(t)\\ \tilde\phi=\tilde\phi(t)}}\partial_t\phi(x,t)+\frac{\delta L[\phi,\tilde\phi]}{\delta\tilde\phi(x)}\biggr\vert_{\substack{\phi=\phi(t)\\ \tilde\phi=\tilde\phi(t)}}\partial_t\tilde\phi(x,t)\right]
\end{equation} 
 to get
 \begin{equation}
 V_0[\eta]=V_0[\phi^*]+L[\eta,\tilde\eta]-L[\phi^*,\tilde\phi^*]+\int_{-\infty}^0 d\tau\,\int_\Lambda dx\,\tilde\pi(x,\tau)\partial_\tau\tilde\phi(x,\tau)\label{qp2}
 \end{equation}
 where the fields $\tilde\eta$ and $\tilde\phi^*$ are the canonical transformed $\eta$ and $\phi^*$ respectively. 
 
At this point, it could look like that we have not gained too much because we still have to do time integral to get the quasipotential. However, we have the possibility to design a convenient form for $(\tilde\phi,\tilde\pi)$ so that the integral in eq. (\ref{action2}) can be done. Therefore, our next step is to look for the necessary assumptions to find such optimal transformation. 
 
 \subsection*{III.2. $V_0$'s convenient form}
 
 In the transformed variables $(\tilde\phi,\tilde\pi)$ we can define its quasipotential by eq. (\ref{action2}), $\tilde V_0[\tilde\phi]$. Therefore 
 \begin{equation}
 \tilde\pi(x)\vert_{\tilde T}=\frac{\delta\tilde V_0[\tilde\phi]}{\delta\tilde\phi(x)}\biggr\vert_{\tilde T}\label{pit}
 \end{equation}
 where $\tilde T$ represent any pair $(\tilde\phi(x,t),\tilde\pi(x,t))$ that are solution of the Hamilton's equations (\ref{Heq}) with the canonical transformed hamiltonian: $\tilde H(\tilde\phi,\tilde\pi)=H(\phi,\pi)$ and the corresponding boundary conditions. This implies two relevant properties:
 \begin{itemize}
   \item  There exists a functional relation between the paths $\phi(t)$ and $\tilde\phi(t)$ solutions of the respective Hamilton's equation of motion:
\begin{equation}
\phi(x,t)=\phi[\tilde\phi(t);x] \label{p1}
\end{equation}
This can be seen by restricting the canonical transformation (\ref{ct}) to the trajectories and using eq. (\ref{pit}):
\begin{equation}
\frac{\delta\tilde V_0[\tilde\phi]}{\delta\tilde\phi(x)}\biggr\vert_{\tilde T}=-\frac{\delta L[\phi,\tilde\phi]}{\delta\tilde\phi(x)}\biggr\vert_{T,\tilde T}\label{B1}
\end{equation}
Therefore from this implicit equation we assume that eq.(\ref{p1}) exists at each $t$ from the original and transformed Hamiltonian trajectories. 
\item  $V_0$ in eq. (\ref{action2}) can be written:
 \begin{equation}
 V_0[\eta]=V_0[\phi^*]+L[\eta,\tilde\eta]-L[\phi^*,\tilde\phi^*]+\tilde V_0[\tilde\eta]-\tilde V_0[\tilde\phi^*]\label{qp3} 
 \end{equation}
Observe from eq.(\ref{qp3}) that the quasipotential is linearly related to the generator of the canonical transformation.  This relation permits us to explain the existence of the extremal property (\ref{ss3}) observed by Derrida et al. in Ref.\cite{Derrida} (let us mention that a similar property was observed by Bertini and co-workers when studying the quasipotential associated with a model of heat flow \cite{Bertini1}):  from eq.(\ref{qp3}) we define  $G[\phi,\tilde\phi]=V_0[\phi^*]+L[\phi,\tilde\phi]+\tilde V_0[\tilde\phi]-\tilde V_0[\tilde\phi]$. Then, the condition $\delta G[\phi,\tilde\phi]/\delta\tilde\phi(x)=0$ on $T,\tilde T$, is just equation (\ref{B1}) that relates the variables $\phi$ with the transformed ones $\tilde\phi$ and therefore $\phi(x)=\phi[\tilde\phi;x]$ is the extremal solution of (\ref{ss3}). This result implies that the auxilary field defined in Derrida et al. paper \cite{Derrida} is just the canonical transformed $\phi$-field. 
 \end{itemize}

These two properties allow us to get a more convenient form to compute the quasipotential algebraically.
Let us define the {\it restricted} transformation $\tilde L$ by substituting $\phi$ by its relation with $\tilde\phi$ along $T$ in (\ref{p1}):
\begin{equation}
\tilde L[\tilde\phi]=L[\phi[\tilde\phi],\tilde\phi]
\end{equation}
then
\begin{equation}
\frac{\delta\tilde L[\tilde\phi]}{\delta\tilde\phi(x)}=-\frac{\delta\tilde V_0[\tilde\phi]}{\delta\tilde\phi(x)}+\int_{\Lambda}dy\,A[\phi[\tilde\phi],\tilde\phi;y]K[\tilde\phi;y,x]\label{der}
\end{equation}
where
\begin{equation}
K[\tilde\phi;x,y]=\frac{\delta\phi[\tilde\phi;x]}{\delta\tilde\phi(y)}\quad ,\quad  A[\phi,\tilde\phi;x]=\frac{\delta L[\phi,\tilde\phi]}{\delta\phi(x)}\label{K}
\end{equation}

Assuming that $\tilde L[\tilde\phi]$ exists (we will address this issue below), we know that (see Appendix 1):
\begin{eqnarray}
\tilde L[\tilde\eta]&=&\tilde L[\tilde\phi^*]-\tilde V_0[\tilde\eta]+\tilde V_0[\tilde\phi^*]\nonumber\\
&+&\int_0^1d\lambda\,\int_\Lambda dx\,(\tilde\eta(x)-\tilde\phi^*(x))\int_\Lambda dy\,A[\phi[\tilde\phi(\lambda)],\tilde\phi(\lambda);y]K[\tilde\phi(\lambda),y,x]
\end{eqnarray}
where
\begin{equation}
\tilde\phi(x,\lambda)=\tilde\phi^*(x)+\lambda(\tilde\eta(x)-\tilde\phi^*(x))
\end{equation}
and  the quasipotential (\ref{qp3}) can be  written on its {\it convenient form}:
\begin{eqnarray}
 V_0[\eta]&=&V_0[\phi^*]\nonumber\\&+&\int_{0}^1d\lambda\int_{\Lambda}dx\,(\tilde\eta(x)-\tilde\phi^*(x))\int_{\Lambda}dy\,A[\phi[\tilde\phi(\lambda)],\tilde\phi(\lambda);y]K[\tilde\phi(\lambda);y,x]
  \label{qp4}
 \end{eqnarray}
We see that the time integral in (\ref{action2}) has been ``deformed'' by a straight path connecting the stationary state $(\tilde\phi^*,0)$ and the target state $(\tilde\eta,\tilde\pi)$. Moreover, to get $V_0$, we do not need to know the full canonical transformation but the functionals $A$ and $K$ defined in eq.(\ref{K}). In expression (\ref{qp4}) is hidden a practical problem: for a given canonical transformation $L$, we get easily the functional $A$ (it is just a functional derivative of $L$), but we cannot obtain the functional relation between $\phi$ and $\tilde\phi$ (eq. (\ref{p1})) because we should solve the Hamilton equations explicitly and get the paths to build the map. In conclusion, it seems that we are stuck with the same problem that we initially had. We cannot follow the natural but somehow impossible scheme of deriving the full canonical transformation and afterward get the functionals $A$ and $\phi[\tilde\phi;x]$.

However,  the relation (\ref{qp4}) can be useful if we change our point of view. Let us assume that the functionals $A$ and $\phi[\tilde\phi;x]$ are given: what are their conditions to guarantee that they come from a well-defined canonical transformation? That would happens whenever (1) $A$ fulfills the conditions for the existence of $L$ and therefore for $\tilde L$, and (2)  $\phi(x,t)=\phi[\tilde\phi(t);x]$ is compatible with Hamilton's equations of motion for $\phi$ and $\tilde\phi$. Therefore, our starting point will initially assume a particular family of functional forms for $A$ and $\phi[\tilde\phi;x]$. Then, we will apply those minimal conditions to determine their detailed functional structure and select the form of the functionals defining the Langevin equation. We will show that this strategy completely determines the functional $A$ and $\phi[\tilde\phi;x]$. Moreover, we also can do all those steps in a systematic algebraic manner. 

\subsection*{III.3. Minimal Conditions on $A$ and $\phi[\tilde\phi;x]$ functionals} 
 
We know that giving the two functionals  $A[\phi,\tilde\phi;x]$ and $B[\phi,\tilde\phi;x]$ in eq.(\ref{ct}), they may considered as the first derivatives of $L[\phi,\tilde\phi]$ if and only if  their doubled crossed derivatives  are independent on the applied order.  That is: 
\begin{eqnarray}
\text{(C1):   }\frac{\delta A[\phi,\tilde\phi;x]}{\delta\phi(y)}&=&\frac{\delta A[\phi,\tilde\phi;y]}{\delta\phi(x)}\nonumber\\
\text{(C2):   }\frac{\delta A[\phi,\tilde\phi;x]}{\delta\tilde\phi(y)}&=&\frac{\delta B[\phi,\tilde\phi;y]}{\delta\phi(x)}\nonumber\\
\text{(C3):   }\frac{\delta B[\phi,\tilde\phi;x]}{\delta\tilde\phi(y)}&=&\frac{\delta B[\phi,\tilde\phi;y]}{\delta\tilde\phi(x)}
\end{eqnarray}
By other hand, equation (\ref{B1}) implies that $B[\phi,\tilde\phi;x]$ should be at least of the form:
\begin{equation}
B[\phi,\tilde\phi;x]=\frac{\delta\tilde V_0[\tilde\phi]}{\delta\tilde\phi(x)}+\int_\Lambda dy\,(\phi(y)-\phi[\tilde\phi;y])B_{\perp}[\phi,\tilde\phi;x,y]
\end{equation} 
Using this expression for $B$ we can write the conditions {\it (C2)} and {\it (C3)} in the following form:
\begin{equation}
\text{(C2'):   }\frac{\delta A[\phi,\tilde\phi;x]}{\delta\tilde\phi(y)}=B_{\perp}[\phi,\tilde\phi;y,x]+\int_\Lambda dz (\phi(z)-\phi[\tilde\phi;z])\frac{\delta B_{\perp}[\phi,\tilde\phi;y,z]}{\delta\phi(x)}
\end{equation}
\begin{eqnarray}
\text{(C3'):   }&&\int_{\Lambda}dz\,\left[ K[\tilde\phi;z,y]B_{\perp}[\phi,\tilde\phi;x,z]- K[\tilde\phi;z,x]B_{\perp}[\phi,\tilde\phi;y,z]\right]\nonumber\\
&=&\int_{\Lambda}dz\, (\phi(z)-\phi[\tilde\phi;z])\left[\frac{\delta B_{\perp}[\phi,\tilde\phi;x,z]}{\delta\tilde\phi(y)}-\frac{\delta B_{\perp}[\phi,\tilde\phi;y,z]}{\delta\tilde\phi(x)}\right]
\end{eqnarray}
We can get a set of {\it necessary} conditions for the existence of $L$ if we restrict them to the trajectory $T$ where $\phi(x)=\phi[\tilde\phi;x]$:
\begin{eqnarray}
\text{(C1T):   }\frac{\delta A[\phi,\tilde\phi;x]}{\delta\phi(y)}\biggr\vert_{\phi=\phi[\tilde\phi]}&=&\frac{\delta A[\phi,\tilde\phi;y]}{\delta\phi(x)}\biggr\vert_{\phi=\phi[\tilde\phi]}\label{C1T}\\
\text{(C2T):   }\frac{\delta A[\phi,\tilde\phi;x]}{\delta\tilde\phi(y)}\biggr\vert_{\phi=\phi[\tilde\phi]}&=&B_{\perp}[\phi[\tilde\phi],\tilde\phi;y,x]\label{C2T}\\
\text{(C3T):   }\int_{\Lambda}dz\,\biggl[K[\tilde\phi;z,y]\frac{\delta A[\phi,\tilde\phi;z]}{\delta\tilde\phi(x)}\biggr\vert_{\phi=\phi[\tilde\phi]}&-&K[\tilde\phi;z,x]\frac{\delta A[\phi,\tilde\phi;z]}{\delta\tilde\phi(y)}\biggr\vert_{\phi=\phi[\tilde\phi]}\biggr]=0\label{C3T}
\end{eqnarray}

Please observe that the conditions (C1T) and (C3T) only depend on the functionals $A$ and $K$, the same ones that we need to get $V_0$ in equation (\ref{qp4}). We are interested in obtaining $V_0$ and, therefore, we will use only (C1T) and (C3T).  (C2T) become just a property that is of no use for our practical purposes. It is out of the scope of this work to attempt to rigorously prove the sufficient conditions (C1), (C2), and (C3) assuming (C1T), (C2T), and (C3T).

The condition on the $\phi[\tilde\phi;x]$ functional is obtained by using the Hamilton's equations of motion.
We know that $\phi$  follow a path that is solution of the Hamilton's equations (\ref{Heq}). Let us rewrite such equations using the canonical transformation (\ref{ct}) and substituting $\phi(x)$ by $\phi[\tilde\phi;x]$ and $\pi(x)$ by $A[\phi,\tilde\phi;x]$:
\begin{equation}
\partial_t\phi[\tilde\phi(t);x]=\tilde R_1[\tilde\phi(t);x]\quad,\quad \partial_t A[\phi[\tilde\phi(t)],\tilde\phi(t);x]=\tilde R_2[\tilde\phi(t);x]\label{e1}
\end{equation}
where
\begin{eqnarray}
R_1[\phi,\tilde\phi;x]&=&\frac{\delta H[\phi,\pi]}{\delta\pi(x)}\biggr\vert_{ \pi=A[\phi,\tilde\phi;x]}\nonumber\\
R_2[\phi,\tilde\phi;x]&=&-\frac{\delta H[\phi,\pi]}{\delta\phi(x)}\biggr\vert_{\pi=A[\phi,\tilde\phi;x]}\label{RR11}
\end{eqnarray}
and $\tilde R_{1,2}[\tilde\phi;x]=R_{1,2}[\phi[\tilde\phi],\tilde\phi;x]$.
We can now expand the time derivatives and we get:
\begin{eqnarray}
\int_{\Lambda}dy\,\frac{\delta\phi[\tilde\phi;x]}{\delta\tilde\phi(y)}\biggr\vert_{\tilde\phi(t)}\,\partial_t\tilde\phi(y,t)&=&\tilde R_1[\tilde\phi(t);x]\nonumber\\
\int_{\Lambda}dy\,\left[\frac{\delta A[\phi,\tilde\phi;x] }{\delta\phi(y)}\biggr\vert_{\substack{\phi=\phi[\tilde\phi(t)]\\ \tilde\phi=\tilde\phi(t)}}\,\tilde R_1[\tilde\phi(t);y]+\frac{\delta A[\phi,\tilde\phi;x] }{\delta\tilde\phi(y)}\biggr\vert_{\substack{\phi=\phi[\tilde\phi(t)]\\ \tilde\phi=\tilde\phi(t)}}\,\partial_t\tilde\phi(y,t)\right]&=&\tilde R_2[\tilde\phi(t);x]
\end{eqnarray}
These equations are combined to disregard their dependence on $\partial_t\tilde\phi$:
\begin{equation}
\int_\Lambda dy\int_{\Lambda}dz\,\frac{\delta A[\phi,\tilde\phi;x] }{\delta\tilde\phi(y)}\biggr\vert_{\phi=\phi[\tilde\phi]}\tilde R_1[\tilde\phi;z]K^{-1}[\tilde\phi;y,z]=\tilde R_2[\tilde\phi;x]-\int_\Lambda dy\,\frac{\delta A[\phi,\tilde\phi;x] }{\delta\phi(y)}\biggr\vert_{\phi=\phi[\tilde\phi]}\tilde R_1[\tilde\phi;y]\label{e2}
\end{equation}
where
\begin{equation}
 \int_{\Lambda}dy\,K[\tilde\phi;x,y]K^{-1}[\tilde\phi;y,z]=\delta(x-z)
\end{equation}
Observe that we have dropped out the time dependence, considering that this relation for the functional $\phi[\tilde\phi]$  holds along each point in the path. We can get a more convenient expression where $K^{-1}$ disappears by integrating both sides by $\int_\Lambda dx K[\tilde\phi;x,z]$ and using the (C3T) property above:
\begin{eqnarray}
(EM):&&\int_\Lambda dx\,\biggl[K[\tilde\phi;x,y]\tilde R_2[\tilde\phi;x]-\nonumber\\
&&\tilde R_1[\tilde\phi;x]\bigg(\frac{\delta A[\phi,\tilde\phi;x] }{\delta\tilde\phi(y)}\biggr\vert_{\phi=\phi[\tilde\phi]}+\int_\Lambda dz\,K[\tilde\phi;z,y] \frac{\delta A[\phi,\tilde\phi;z] }{\delta\phi(x)}\biggr\vert_{\phi=\phi[\tilde\phi]}\bigg)\biggr]=0\label{EM}
\end{eqnarray}
(EM) is the minimal condition on the functionals $A$ and $\phi[\tilde\phi;x]$ to make them compatible with the Hamilton's equations of motion.

Therefore, the set of necessary conditions over the  $A$ and $\phi[\tilde\phi]$ functionals that we are to consider here are (C1T) (eq.\ref{C1T}), (C2T) (eq.\ref{C2T}) and (EM) (eq.(\ref{EM})).  
As we already said, to go further, we will propose some functional forms for both functionals. Let us do a particular example to illustrate how these conditions are enough to define the functionals $A$ and  $\phi[\tilde\phi]$ and how we get the corresponding quasipotential.

\subsection*{III.4. The full method at work: an example}
Let us choose as an example the one-dimensional diffusive model defined in section II by the Langevin equation (\ref{Dmodel1}) with the determinist part given by (\ref{Dmodel2}). Observe that the functions $D(\phi)$ and $\chi(\phi)$ are unspecified. The initial step for this method is to choose a concrete family of functionals for $A$ and $\phi[\tilde\phi;x]$. In this example we elect the simplest functional forms:
\begin{eqnarray}
A[\phi,\tilde\phi;x]&=&a(\phi(x),\tilde\phi(x),\phi^*(x))\nonumber\\
 \phi(x)&=&f(\tilde\phi(x),\phi^*(x))\label{ch0}
\end{eqnarray}
where $a$ and $f$ are functions to be determined and $\phi^*(x)$ is the deterministic stationary state.  We have included an explicit dependence on the stationary state because we know that $\pi^*(x)=A[\phi^*,\tilde\phi^*;x]=0$ and such degree of freedom is necessary. We also should take into account that, at the fix boundaries (if any) $\pi(x)=A[\phi,\tilde\phi;x]=0\quad\forall x\in\partial\Lambda$. Let us remark that eq. (\ref{ch0}) is an arbitrary choice. We could have included any derivative of $\phi(x)$ or $\tilde\phi(x)$ on the arguments of both functions and/or any other more complex functional non-local structure. We will comment below about the choices that we are able to handle. Let us stress that a priori, there is no guarantee that a given family of functionals is going to be compatible with the set of conditions (C1T), (C3T), and (EM) above. The election (\ref{ch0}) implies:
\begin{eqnarray}
\frac{\delta A[\phi,\tilde\phi;x]}{\delta\phi(y)}&=& a^{(1,0,0)}(u_0,v_0,u^*_0)\delta(x-y)\nonumber\\
\frac{\delta A[\phi,\tilde\phi;x]}{\delta\tilde\phi(y)}&=& a^{(0,1,0)}(u_0,v_0,u^*_0) \delta(x-y)\nonumber\\
K[\tilde\phi;x,y]&=&\frac{\delta\phi[\tilde\phi;x]}{\delta\tilde\phi(y)}=f^{(1,0)}(v_0,u^*_0)\delta(x-y)\label{ch1}
\end{eqnarray}
where $a^{(n,m,l)}(u_0,v_0,u^*_0)\equiv\partial^{n+m+l}a(n,m,l))/\partial u^n_0\partial v^m_0\partial u^{*l}_0$. In order to simplify notation we use the convention that $u_k\leftrightarrow\phi^{(k)}(x)=d^k\phi(x)/dx^k$, $v_k\leftrightarrow\tilde\phi^{(k)}(x)$ and $u^*_k\leftrightarrow\phi^{*(k)}(x)$.
The minimal conditions to have a well defined canonical transformations are, in this case, given by:
\begin{itemize}
\item{(C1T):} $\delta A[\phi,\tilde\phi;x]/\delta\phi(y)$ is symmetric under the $x$, $y$ exchange. Therefore this condition is always fulfilled for this choice.
\item{(C3T):}  $\delta A[\phi,\tilde\phi;x]/\delta\tilde\phi(y)$ is also symmetric under the $x$, $y$ exchange. Therefore this condition is always fulfilled for this choice.
\item{(EM):} After we substitute eqs.(\ref{ch1}) into the condicion (\ref{EM}) we get:
\begin{eqnarray}
f^{(1,0)}(v_0,u^*_0)\tilde R_2[\tilde\phi;x]&&=\nonumber\\
\biggl[a^{(0,1,0)}(u_0,v_0,u^*_0)&&+a^{(1,0,0)}(u_0,v_0,u^*_0)f^{(1,0)}(v_0,u^*_0)\biggr]\tilde R_1[\tilde\phi;x]
\end{eqnarray}
where  $u_0=f(v_0,u^*_0)$.
Let us remind that $\tilde R_{1,2}[\tilde\phi;x]=R_{1,2}[\phi[\tilde\phi],\tilde\phi;x]$ and $R_{1,2}[\phi,\tilde\phi;x]$ are defined by eq.(\ref{RR11}) with the hamiltonian (\ref{Honedif}). They have the form:
\begin{eqnarray}
R_1[\phi,\tilde\phi;x]&=&-\frac{d}{dx}\left[-D(\phi(x))\phi'(x)+(E+\frac{d a(x)}{dx}) \chi(\phi(x))\right]\nonumber\\
R_2[\phi,\tilde\phi;x]&=&-\chi'(\phi(x))\frac{d a(x)}{dx}\left(E+\frac{1}{2}\frac{d a(x)}{dx}\right)-D(\phi(x))\frac{d^2a(x)}{dx^2}\label{R1R2dif}
\end{eqnarray}
where $a(x)\equiv a(\phi(x),\tilde\phi(x);\phi^*(x))$.
\end{itemize}

After doing all the functional substitutions and spatial derivatives, the unique non-trivial condition (EM) is  a differential equation of the form:
\begin{equation}
c_3(v_0,u^*_0)v_2+c_2(v_0,u^*_0,u^*_1)v^2_1+c_1(v_0,u^*_0,u^*_1)v_1+c_0(v_0,u^*_0,u^*_1)=0\label{poly}
\end{equation}
where we are using our simplified notation $v_k\leftrightarrow\tilde\phi^{(k)}(x)=d^k\tilde\phi(x)/dx^k$.
For instance, 
\begin{eqnarray}
&&c_3(v_0,u^*_0)=\left(f^{(1,0)}(v_0,u^*_0) a^{(1,0,0)}(u_0,v_0,u^*_0)+a^{(0,1,0)}(u_0,v_0,u^*_0)\right) \nonumber\\
&&\left(\chi(u_0) \left(f^{(1,0)}(v_0,u^*_0) a^{(1,0,0)}(u_0,v_0,u^*_0)+a^{(0,1,0)}(u_0,v_0,u^*_0)\right)-2 D(u_0) f^{(1,0)}(v_0,u^*_0)\right)\label{poly3}
\end{eqnarray}
where  we remind that $u_0=f(v_0,u^*_0)$. $c_0$, $c_1$ and $c_2$ have large similar expressions that we do not show for simplicity. 
We know that we are assuming that the canonical transformation relates $\phi(x)$ and $\tilde\phi(x)$ through our choice $\phi(x)=f(\tilde\phi(x),\phi^*(x))$ {\it for any} $\phi(x)$ (compatible with the boundary conditions). However, the differential equation (\ref{poly}) is in itself an apparent contradiction because its existence would imply that we can canonically transform only its finite number of solutions. This is not the proper interpretation. Equation (\ref{poly}) is not a differential equation because it shold be correct {\it for any} $\phi(x)$ ($\tilde\phi(x)$) function. In fact,  for any given and arbitrary point $x\in\Lambda$,  the field $\tilde\phi(x)$ and its derivatives at such point,$v_0, v_1, \ldots v_k$, have arbitrary values by  varying $\phi(x)$. 
Therefore we have to consider eq.(\ref{poly}) as a polynomial formed by independent variables $v_0, v_1,\ldots$. The polynomial can only be zero for any of $v_k$-values when each coefficient $c_0, c_1,\ldots$ is identically equal to zero. This give us a set of conditions that fix the functional forms for $f(v_0,u^*_0)$ and $a(u_0,v_0,u^*_0)$. We see that the complexity of the coefficients decreases with the derivative order. Therefore it is convenient to extract first the information from $c_3(v_0,u^*_0)=0$, substitute any condition obtained into the polynomial expression (\ref{poly}) and afterwards we study $c_2(v_0,u^*_0,u^*_1)=0$, and so on.

The condition $c_3(v_0,u^*_0)=0$ have two factors that should be studied separately:
\begin{itemize}
\item {1. $f^{(1,0)}(v_0,u^*_0) a^{(1,0,0)}(u_0,v_0,u^*_0)+a^{(0,1,0)}(u_0,v_0,u^*_0)$=0}

This condition implies the functional relation:
\begin{equation}
f^{(1,0)}(v_0,u^*_0)=-\frac{\displaystyle a^{(0,1,0)}(f(v_0,u^*_0),v_0,u^*_0) }{\displaystyle  a^{(1,0,0)}(f(v_0,u^*_0),v_0,u^*_0)}
\end{equation}
When we use this relation back into eq.(\ref{poly}) we get
\begin{equation}
f^{(0,1)}(v_0,u^*_0)=-\frac{\displaystyle a^{(0,0,1)}(f(v_0,u^*_0),v_0,u^*_0) }{\displaystyle  a^{(1,0,0)}(f(v_0,u^*_0),v_0,u^*_0)}
\end{equation}
Both relations imply that:
\begin{equation}
\frac{d a(\phi(x),\tilde\phi(x),\phi^*(x))}{dx}=0\quad\Rightarrow\quad \pi(x)=0 \quad \forall x\in\Lambda
\end{equation}
That is a particular trivial solution of our conditions corresponding to being at the stationary state.

\item{2. $\chi(u_0) \left(f^{(1,0)}(v_0,u^*_0) a^{(1,0,0)}(u_0,v_0,u^*_0)+a^{(0,1,0)}(u_0,v_0,u^*_0)\right)-2 D(u_0) f^{(1,0)}(v_0,u^*_0)=0$ }

This relation can be written as:
\begin{equation}
\frac{\partial}{\partial v_0}\left[a(f(v_0,u^*_0),v_0,u^*_0)-\int^{f(v_0,u^*_0)}du\frac{2D(u)}{\chi(u)}\right]=0
\end{equation}
that implies
\begin{equation}
a(f(v_0,u^*_0),v_0,u^*_0)=\int^{f(v_0,u^*_0)}du\frac{2D(u)}{\chi(u)}+h(u^*_0)
\end{equation}
where $h(u)$ is a function to be determined. Observe that $a(u_0,v_0,u^*_0)$ depends explicitly on $v_0$ just through $u_0=f(v_0,u^*_0)$. We use the fact that $a(u^*_0,u^*_0)=0$ to determine $h(u^*_0)$. The final expresion is then
\begin{equation}
a(u_0,u^*_0)=\int_{u^*_0}^{u_0}du\frac{2D(u)}{\chi(u)}
\end{equation}
This expression for $a$ reduces the polynomial (\ref{poly}) to:
\begin{equation}
\frac{2 u^*_1 f^{(1,0)}(v_0,u^*_0) (u^*_1 D(u^*_0)-E \chi(u^*_0)) \left(D(u^*_0) \chi'(u_0)-D(u_0) \chi'(u^*_0)\right)}{\chi(u^*_0)^2}=0
\end{equation}
We see that the unique non-trivial condition (the one that gives us a quasipotential different from the equilibrium one) is obtained when
\begin{equation}
 D(u^*_0) \chi'(u_0)=D(u_0) \chi'(u^*_0)\quad\Rightarrow\quad D(u)=c\chi'(u)
\end{equation}
The diffusive model having this relation between the diffusion and the mobility is called {\it zero-range model}.
\end{itemize}
Let us remark that from the conditions (C1T), (C3T) and (EM) we have determined completely $a(u_0,v_0,u^*_0)$ and a restricted family of diffusive systems (the zero-range model) compatible with the initial assumptions on the functionals $A$ and $\phi[\tilde\phi;x]$. However, it seems that we do not know the precise form of $f(v_0,u^*_0)$. We will see that, in this case, the quasipotential can be computed for {\it any} well defined $f$.

Once we know the form of $a$ and the models that can be described by this functional forms we compute the quasipotential by using eq.(\ref{qp4}) that in this case can be written:
\begin{equation}
V_0[\eta]=V_0[\phi^*]+\int_0^1d\lambda\int_\Lambda dx\,(\tilde\eta(x)-\phi^*(x))\int_\Lambda dy\,a(\phi[\tilde\phi(\lambda);y])\frac{\delta\phi[\tilde\phi(\lambda);y]}{\delta\tilde\phi(x,\lambda)}
\end{equation}
where  $\phi[\tilde\phi;x]=f(\tilde\phi(x);\phi^*(x))$ and $\tilde\phi(x;\lambda)=\phi^*(x)+\lambda(\tilde\eta(x)-\phi^*(x))$. We do easily the functional derivative: $\delta\phi[\tilde\phi(\lambda);y]/\delta\tilde\phi(x,\lambda)=\partial f(v_0;\phi^*(x))/\partial v_0\vert_{v_0=\tilde\phi(x;\lambda)}\delta(x-y)$. Then
\begin{eqnarray}
V_0[\eta]&=&V_0[\phi^*]+2c\int_\Lambda dx\int_0^1d\lambda\,(\tilde\eta(x)-\phi^*(x))\frac{\partial f(v_0;\phi^*(x))}{\partial v_0}\biggr\vert_{v_0=\tilde\phi(x;\lambda)}\nonumber\\
&&\log\frac{\chi(f(\tilde\phi(x;\lambda);\phi^*(x)))}{\chi(\phi^*(x))}
\end{eqnarray}
where we have used eq.(\ref{qp4}) and $D(u)=c\chi'(u)$. We make the change of variables $\lambda\rightarrow f(\tilde\phi(x;\lambda);\phi^*(x))$ at each $x$-value and we get:
{\boldmath
\begin{equation}
V_0[\eta]=V_0[\phi^*]+2c\int_\Lambda dx\int_{\phi^*(x)}^{\eta(x)} dw\log\frac{\chi(w)}{\chi(\phi^*(x))}\label{quasii}
\end{equation}}
Where $\phi^*(x)$ is the stationary state that in this case it is solution of
\begin{equation} 
c\frac{d^2}{dx^2}\chi(\phi^*(x))=E\frac{d}{dx}\chi(\phi^*(x))
\end{equation}
For periodic boundary conditions the solution is a constant: $\phi^*(x)=\phi^*$  fixed by the initial condition:
\begin{equation}
\frac{1}{\vert\Lambda\vert}\int_\Lambda dx \,\phi(x,0)=\phi^*
\end{equation} 
and the current is $J=-E\chi(\phi^*)$. For fixed boundary conditions at $x=0$ and $x=1$, $\phi(0,t)=\phi_0$ and $\phi(1,t)=\phi_1$ respectively, then
\begin{equation}
\chi(\phi^*(x))=\frac{\chi_1\left(e^{\bar E x}-1\right)+\chi_0\left(e^{\bar E}-e^{\bar E x}\right)}{e^{\bar E}-1}
\end{equation}
where $\bar E=E/c$ and $\chi_{0,1}=\chi(\phi_{0,1})$. The current is given by
\begin{equation}
J=E\frac{\chi_1-\chi_0 e^{\bar E}}{e^{\bar E}-1}
\end{equation}
Observe that equilibrium is obtained when $J=0$, that is, when $\chi_1=\chi_0 e^{\bar E}$ and the equilibrium stationary profile is given by $\chi(\phi_{eq}^*)=\chi_0 e^{\bar E x}$. The quasipotential becomes, in this case, the {\it equilibrium potential} that we could have derived directly from the Einstein relation (\ref{er}) together with $D(\phi)=c\chi'(\phi)$. 
We see that here is no difference between equilibrium and non-equilibrium in the mathematical structure of the quasipotential. This happens uniquely for this particular model (see Ref.\cite{Bertini}). 

In (2001), Bertini et al. \cite{Bertini0} obtained the quasipotential for $c=1$. We recover their expression by assuming: $\chi(u)=S(u)$, $D(u)=cS'(u)$ and making the change of variables in eq.(\ref{quasii}): $w=sZ'(s)/Z(s)$ where $Z(s)$ is such that $S(w)=s$:
{\boldmath
\begin{equation}
V_0[\eta]=V_0[\phi^*]+2c\int_\Lambda dx\left[\eta(x)\log\frac{S(\eta(x))}{S(\phi^*(x))}-\log\frac{Z(S(\eta(x)))}{Z(S(\phi^*(x)))} \right]
\end{equation}}

In this explicit example, we have shown how our method works. First, we gave a Langevin equation. Second we defined a particular family of functionals for $A[\phi,\tilde\phi;x]$ and $\phi[\tilde\phi;x]$. Third, we used the conditions (C1T), (C3T), and (EM) to fix the detailed structure of the unknown functionals. We were forced during that derivation to introduce some conditions on the Langevin equation: $D(\phi)=c\chi'(\phi)$. Finally, we computed the quasipotential.  We'll see that this scheme applies to all examples we have worked out in the paper. Let us mention that there is no {\it a priori} guarantee that the scheme should always give a solution. Sometimes we have found that some conditions are never fulfilled for a Langevin equation and a given family of functionals $A$ and $\phi[\tilde\phi;x]$. 

\section{IV. Generic Functional form for $A$ and $\phi[\tilde\phi]$ in 1-d systems}

In this section we restrict ourselves to one-dimensional models and we generalize the choice of the $A$ and $\phi[\tilde\phi]$ functionals we have used in the example above. We study the mathematical form of the conditions (C1T), (C3T) and (EM) for such generalization. We'll see that we can algebraically handle only a subfamily of such general choice. Finally we will discuss some properties of such subfamily. 

\subsection*{IV.1.  A family of functional forms}

Let us assume that the local functional $A[\tilde\phi;x]$ is of the form:
\begin{equation}
A[\phi,\tilde\phi;x]=a(\phi(x),\phi'(x),\ldots,\phi^{(n)}(x),\tilde\phi(x),\tilde\phi'(x),\ldots,\tilde\phi^{(m)}(x),\phi^*(x))\label{a}
\end{equation}
for given arbitrary integer values of $m,n\geq 0$. Where, as we already saw in Section II.4, we have included an explicit dependence on the stationary state because we know that $\pi^*(x)=A[\phi^*,\tilde\phi^*;x]=0$ and such degree of freedom maybe necessary in some cases. Similarly, at the fix boundaries (if any) $\pi(x)=A[\phi,\tilde\phi;x]=0\quad\forall x\in\partial\Lambda$. Both properties should be taken into account when defining the $a$ function.

Our second choice is to take a generic form for the functional $\phi[\tilde\phi]$. We consider that there exists an implicit relation of the form:
\begin{equation}
\phi^{(l)}(x)=f(\phi(x),\phi'(x),\ldots,\phi^{(l-1)}(x),\tilde\phi(x),\tilde\phi'(x),\ldots,\tilde\phi^{(s)}(x),\phi^*(x))\label{f}
\end{equation}
for given arbitrary values of $l>0$ and $s\geq 0$. We also study the case 
\begin{equation}
\phi(x)=f(\tilde\phi(x),\tilde\phi'(x),\ldots,\tilde\phi^{(s)}(x),\phi^*(x))\label{ff}
\end{equation}

We choose this form for the functional relation between $\phi$ and $\tilde\phi$ because it is ``algebraically simple'' and it may capture the complexity of a non-equilibrium system as it was shown by the work of Derrida et al. \cite{Derrida} that we briefly explained in the introduction. Nevertheless, one may attempt some other more complex possibilities, but, as we will see, with our election, we are already at the edge of today's mathematics  that permits us to get  explicit solutions to our problem.  
Observe that in eq.(\ref{f}) we are looking for a non-linear differential equation of order $s$ whose solution gives us the relation between $\tilde\phi$ and $\phi$. Such differential equations have typically $s$ arbitrary constants. Some of them can be fixed by the boundary conditions. If there are constants still free after applying the boundary conditions, they should be fixed by looking the ones that minimize the quasipotential. 

\subsection*{IV.2.  Re-writting the conditions (C1T), (C3T) and (EM)}

The  elected local functionals (\ref{a}) and (\ref{f}) permit us re-write with the conditions (CT1), (CT3) and (EM) in a more compact form.
Let us define the differential operators:
\begin{eqnarray}
P_n(x)&=&\sum_{k=0}^n\frac{\partial a}{\partial u_k}\biggr\vert_{\substack{u=\phi \\v=\tilde\phi}}\frac{d^k}{dx^k}\label{P}\\
Q_m(x)&=&\sum_{k=0}^m\frac{\partial a}{\partial v_k}\biggr\vert_{\substack{u=\phi \\v=\tilde\phi}}\frac{d^k}{dx^k}\label{Q}\\
L_l(x)&=&\sum_{k=0}^{l-1}\frac{\partial f}{\partial u_k}\biggr\vert_{\substack{u=\phi \\v=\tilde\phi}}\frac{d^k}{dx^k}-\frac{d^l}{dx^l}\quad l>0\label{L}\\
S_s(x)&=&\sum_{k=0}^s\frac{\partial f}{\partial v_k}\biggr\vert_{\substack{u=\phi \\v=\tilde\phi}}\frac{d^k}{dx^k}\label{S}
\end{eqnarray}
$L_0=-1$. Where $a=a(u_0,u_1,\ldots,u_n,v_0,v_1,\ldots,v_,u^*_0)$ and similarly $f$. Where the convention is that after any derivative of their arguments we should do the substitution $u_k\rightarrow\phi^{(k)}(x)$ and  $v_k\rightarrow\tilde\phi^{(k)}(x)$.

Moreover, we can do a functional derivative with respect to $\tilde\phi(y)$ in both sides of eq.(\ref{f}) or (\ref{ff}) to obtain an equation for the functional $K[\tilde\phi;x,y]=\delta\phi[\tilde\phi;x]/\delta\tilde\phi(y)$:
\begin{equation}
L_l(x)K[\tilde\phi;x,y]+S_s(x)\delta(x-y)=0\label{Ke}
\end{equation}

Let us rewrite the conditions  by using these differential operators:
\begin{itemize}
\item (C1T): Equation (\ref{C1T}) can we written in this case as 
$P_n(x)\delta(x-y)=P_n(y)\delta(x-y)$. 
This happens if and only if the operator $P_n(x)$ is self-adjoint, $P_n^\dag(x)=P_n(x)$ (see Appendix II for a brief reminder about properties and definitions of self-adjoint linear differential operators). We can prove this by observing that by definition and for any $w$ integrable test function
\begin{eqnarray}
&&\int dx w(x)P_n(x)\delta(x-y)=\int dx (P_n^\dag(x) w(x))\delta(x-y)=P_n^\dag(y)w(y)\nonumber\\
&=&\int dx w(x)P_n^\dag(y)\delta(x-y)\Rightarrow P_n(x)\delta(x-y)=P_n^\dag(y)\delta(x-y)
\end{eqnarray}
That is, 
\begin{equation}
(C1T): P_n(x)=P_n^\dag(x)\label{C1T0}
\end{equation}
\item (C3T): Equation (\ref{C3T}) becomes:
\begin{equation}
Q_m^\dag(x)K[\tilde\phi;x,y]=Q_m^\dag(y)K[\tilde\phi;y,x]
\end{equation}
and after using eq.(\ref{Ke}) we find:
\begin{equation}
(C3T): Q_m^\dag(x)(L_l(x))^{-1}S_s(x)=S_s^\dag(x)(L_l^\dag(x))^{-1}Q_m(x)\label{C3T0}
\end{equation}
that is $Q_m^\dag(x)(L_l(x))^{-1}S_s(x)$ should be self-adjoint.

\item (EM): After some trivial algebra we find that (\ref{EM}) can be written:
\begin{equation}
Q_m^\dag(x)\tilde R_1[\tilde\phi;x]=S_s^\dag(x)(L_l^\dag(x))^{-1}\left(P_n(x)\tilde R_1[\tilde\phi;x]-\tilde R_2[\tilde\phi;x]\right)\label{EMv1}
\end{equation}
or, similarly
\begin{equation}
L_l(x)\tilde R_1[\tilde\phi;x]=S_s(x)(Q_m(x))^{-1}\left(P_n(x)\tilde R_1[\tilde\phi;x]-\tilde R_2[\tilde\phi;x]\right)\label{EMv2}
\end{equation}
\end{itemize}

Note that we define the inverse of any differential operator $T(x)$ through its associated Green function:
\begin{equation}
T(x)f(x)=g(x)\Rightarrow f(x)=\int_\Lambda dy\,G(x,y)g(y)\equiv (T(x))^{-1}g(x)
\end{equation}
where $G(x,y)$ is solution of
\begin{equation}
T(x)G(x,y)=\delta(x,y)
\end{equation}

At this point, let us remind our goal: we want to find the functions $a$ and $f$ (given $n$, $m$, $l$ and $s$) such that they fulfill the conditions (C1T), (C3T) and (EM). Inverse differential operators' presence makes it almost impossible to find a systematic algebraic way to get the unknown functions. For instance, we should first find the Green function associated with such a still unknown operator. We know that it can be done systematically for regular boundary value problems (self-adjoint differential operators) once we know the eigenfunctions and eigenvalues of the operator (see, for instance, Ref.\cite{Green}). However, those depend again on the explicit form of the operator. Therefore, from our present knowledge on these issues, it is almost impossible to get algebraically a set of eigenfunctions of our operators $L$, or $Q$ that are unknown functionals of $\tilde\phi$ and its local derivatives.  Consequently, we are going to consider in this paper only situations where there aren't inverse operators in  (C3T) and (EM). There are two possibilities that are compatible with such practical limitations:
\begin{itemize}
\item (a) $l=0$: $L_0(x)=-1$. In this case the operator $L$ is just a constant and the conditions are:
\begin{eqnarray}
\text{(C1T):  }&&P_n(x)=P_n^\dag(x)\nonumber\\
\text{(C3T):  }&&Q_m^\dag(x)S_s(x)=S_s^\dag(x)Q_m(x)\nonumber\\
\text{(EM):   }&&Q_m^\dag(x)\tilde R_1[\tilde\phi;x]=-S_s^\dag(x)\left(P_n(x)\tilde R_1[\tilde\phi;x]-\tilde R_2[\tilde\phi;x]\right)\label{casoa}
\end{eqnarray}
\item (b) $m=0$: $Q_0(x)=\partial a/\partial u_0\vert_{u=\phi,v=\tilde\phi}\neq 0$. In this case the operator $Q$ is just a function and the conditions are written:
\begin{eqnarray}
\text{(C1T):  }&&P_n(x)=P_n^\dag(x)\nonumber\\
\text{(C3T):  }&&S_s(x)Q_0(x)^{-1}L_l^\dag(x)=L_l(x)Q_0(x)^{-1}S_s^\dag(x)\nonumber\\
\text{(EM):   }&&L_l(x)\tilde R_1[\tilde\phi;x]=S_s(x)Q_0(x)^{-1}\left(P_n(x)\tilde R_1[\tilde\phi;x]-\tilde R_2[\tilde\phi;x]\right)\label{casob}
\end{eqnarray}
where we have used eq.(\ref{EMv2}).
\end{itemize}

 In Section V we will systematically consider different scenarios (a) and (b) by applying them to some typical models as the diffusive dynamics and the reaction-diffusion dynamics. 

\subsection*{IV.3. Allowed operators and computation strategy}

We saw that in cases (a) and (b) (Eqs. (\ref{casoa}) and (\ref{casob}) respectively) we should choose a set of values $(n,m,l,s)$ that define the form of operators $P$, $Q$,$L$ and $S$ respectively. From condition (C1T), we know that  $n$ should be even to fulfill the condition that $P$ is self-adjoint. Similarly, (C3T) in case (a) implies $m+s$ should be also even, and in case (b), $s+l$ should also be even. We can do a little better by keeping a trace of the larger derivative of $\tilde\phi(x)$ for each condition. In this case, we should explicitly define the dynamics. We have done such computation for $l=0$ (case (a)) and the Diffusive and the Reaction-Diffusion dynamics defined in section II. We found that for such models the operators should follow the consistency relations:
\begin{equation}
n=0 \Rightarrow m\leq s\quad,\quad n>0\Rightarrow m=s+n
\end{equation}

That is, we can attempt the following cases:
\begin{table}[h]
\begin{center}
\begin{tabular}{| c | c |}
\hline
\multicolumn{2}{| c |}{(n,m,l,s)}\\
\hline
\phantom{ }Case (a) (l=0)\phantom{ }&\phantom{ }Case (b) (m=0)\phantom{ }\\
\hline
{\bf(0,0,0,0)}&{\bf(0,0,1,1)}\\
{\bf(0,1,0,1)}&{\bf(0,0,2,2)}\\
{\bf(0,0,0,2)}&{\bf(0,0,1,3)}\\
{\bf(0,0,0,4)}&{\bf(0,0,3,1)}\\
{\bf(0,1,0,3)}&...\\
{\bf(0,2,0,2)}&{\bf(2,0,1,1)}\\
...&{\bf(2,0,2,0)}\\
\bf{(2,2,0,0)}&...\\
...&\\
\hline
\end{tabular}
\end{center}
\end{table}
where at each column we fix $n=0$ and then $n=2$ and so on.  We have excluded the coincident values between  case (a) and case (b). Once we choose one of these permitted values $(n,m,l,s)$, the method intends to fix the mathematical form of the functions $a$ (\ref{a}) and $f$ (\ref{f}). The main idea is that once we assume the form of $a$ and $f$, the conditions (C1T), (C3T), and (EM) should become identities. In general, these conditions are differential equations. For instance let us assume that $f$ is of the form: $\phi(x)=f(\tilde\phi(x),\tilde\phi'(x),\tilde\phi''(x))$. Let us assume that we can isolate the highest derivative $\tilde\phi''(x)=F_2(\phi(x),\tilde\phi(x),\tilde\phi'(x))$. Then $\tilde\phi^{(3)}(x)=F_3(\phi(x),\phi'(x),\tilde\phi(x),\tilde\phi'(x))$, and in general, $\tilde\phi^{(k)}(x)=F_k(\phi(x),\phi'(x),\ldots,\phi^{(k-2)}(x),\tilde\phi(x),\tilde\phi'(x))$.  On the other hand, $\phi(x)$ is assumed to be an analytic function. Therefore, for any given point in the domain, $\bar x\in\Lambda$, we can reconstruct $\phi(x)$ just by giving all their derivatives at such point. However, $\phi(x)$ is arbitrary, and we are free to choose all the derivatives of $\phi(x)$ at $\bar x$. Therefore, we may reasonably assume that $\tilde\phi^{(k)}(\bar x)$ for $k\geq 2$ may get arbitrary and independent values because of their dependence on the derivatives of $\phi(\bar x)$.  
Having this in mind, we see that the conditions (C1T), (C3T), and (EM) are just polynomials where there are derivatives of $\phi$ and $\tilde\phi$ of a different order. They should be correct {\it for any $\phi$-field} and therefore, {\it for any value of the derivatives of $\tilde\phi(x)$} of degree greater or equal to two. 
Therefore, once we substitute $\phi$ by $f(\tilde\phi,\tilde\phi'(x),\tilde\phi''(x))$ in any of the conditions, we get a polynomial  expression in the derivatives $\tilde\phi^{(n)}$ with $n\geq 2$ whose coefficients are functions with $\tilde\phi(x)$, $\tilde\phi'(x)$ and $\tilde\phi''(x)$ (in this example).  Each of these high order derivatives may have arbitrary values and each of their coefficients should be identically equal to zero. The coefficients that we equal to zero contains, typically derivatives of $f$ and $a$, and they also contain functions that depend on the dynamics. Therefore, $f$ and $a$ may depend on the dynamics, and, sometimes, only a particular dynamics can make zero a coefficient. 

This scheme is done orderly from higher to small order in the polynomial of the derivatives of $\tilde\phi$ for each condition. Once we determine some property of the unknown functions, we include it in the conditions, and we redo the computations to get the remaining high-order derivatives' next coefficient. This method has been applied successfully using algebraic programs like Mathematica. It permits us to do long computations without errors. That is very important because we are dealing with identities, and any small mistake during the algebraic trivial but lengthy evaluation implies that the conditions (C1T), (C3T), and (EM) are never fulfilled.

We should also take into account that $\pi(x)=a(x)$. Therefore it is mandatory that:
\begin{eqnarray}
&&a(\phi^*(x),(\phi^*)'(x),\ldots,(\phi^*)^{(n)}(x),(\tilde\phi)^*(x),(\tilde\phi^*)'(x),\ldots,(\tilde\phi^*)^{(m)}(x);\phi^*(x))=0\nonumber\\
&&a(\phi(x),\phi'(x),\ldots,\phi^{(n)}(x),\tilde\phi(x),\tilde\phi'(x),\ldots,\tilde\phi^{(m)}(x);\phi^*(x))=0\quad \forall x\in\partial\Lambda \label{coco1}
\end{eqnarray}
where $\tilde\phi^*(x)$ is the stationary state in the new variables that is related with the original stationary state $\phi^*(x)$ through eq.(\ref{f}):
\begin{equation}
(\phi^*)^{(l)}(x)=f(\phi^*(x),(\phi^*)'(x),\ldots,(\phi^*)^{(l-1)}(x),\tilde\phi^*(x),(\tilde\phi^*)'(x),\ldots,(\tilde\phi^*)^{(s)}(x);\phi^*(x))\label{f2}
\end{equation}
Observe that the second condition in eq. (\ref{coco1}) only applies when the field's values are fixed at the system's boundaries. 
These conditions helps us to determine $a$ and $f$ functions. 

About the stationary state there are two possibilities: (i) $\tilde\phi^*(x)=\phi^*(x)$ or (ii) $\tilde\phi^*(x)\neq\phi^*(x)$ and this affects to the boundary conditions for the $\tilde\phi(x)$ fields that are necessary to solve the differential equation (\ref{f}). The (i) case is the most convenient (but more restrictive) because implies that both fields have the same boundary conditions $\tilde\phi(x)=\phi(x)\, \forall x\in\partial\Lambda$. Case (ii) needs that eq.(\ref{f2}) to be solved explicitly and then fix some of the $s$ constants by  given the boundary values for $\phi(x)$ and the conditions that $\tilde\phi(x)$  do not evolve at the boundaries: $\tilde R_{1,2}[\tilde\phi;x]=0\, \forall x\in\partial\Lambda$ (see eq.(\ref{e1}) and definitions below it).

\section{V. Quasipotentials for one dimensional Diffusive Models}

We show in this section the quasipotentials obtained by electing some concrete values $(n,m,l,s)$ that define the differential operators on the conditions (C1T), (C3T) and (EM) in eqs.(\ref{casoa}) and (\ref{casob}). We have skept some fine details on their derivations and we refer to the explicit example on Section III.4 to fill the gaps.

\subsection*{\bf (i) {\boldmath$(n,m,l,s)=(0,0,0,0)$}}

This case corresponds to the explicit example we studied in Section III.4 (Zero-range model: {\boldmath$D(\phi)=c\chi'(\phi)$}) where we got its quasipotential. Observe that the corresponding operators (eqs. (\ref{P},\ref{Q},\ref{L},\ref{S})) have the form:
\begin{eqnarray}
P_0(x)&=&a^{(1,0,0)}(u_0,v_0,u^*_0)\nonumber\\
Q_0(x)&=&a^{(0,1,0)}(u_0,v_0,u^*_0)\nonumber\\
L_0(x)&=&-1\nonumber\\
S_0(x)&=&f^{(1,0)}(v_0,u^*_0)
\end{eqnarray} 
and one can check that the general conditions (\ref{casoa}) are reduced to the ones obtained in Section III.4.

\subsection*{\bf (ii) \boldmath$(n,m,l,s)=(0,0,0,2)$}

In this case we choose $a=a(\phi(x),\tilde\phi(x))$ and $\phi(x)=f(\tilde\phi(x),\tilde\phi'(x),\tilde\phi''(x))$. The operators (eqs. (\ref{P},\ref{Q},\ref{L},\ref{S})) have the form:
\begin{eqnarray}
P_0(x)&=& a^{(1,0)}(u_0,v_0)\nonumber\\
Q_0(x)&=& a^{(0,1)}(u_0,v_0)\nonumber\\
L_0(x)&=&-1\nonumber\\
S_2(x)&=&f^{(1,0,0)}(\underline{v})+f^{(0,1,0)}(\underline{v})\frac{d}{dx}+f^{(0,0,1)}(\underline{v})\frac{d^2}{dx^2}
\end{eqnarray} 
where $\underline{v}=(v_0,v_1,v_2)$,  $u_k\equiv\phi^{(k)}(x)$, $v_k\equiv\tilde\phi^{(k)}(x)$. The sufficient conditions given by the cases (a) and (b) (\ref{casoa},\ref{casob}) are equivalent in this case. They are:
\begin{itemize}
\item{\it (C1T):} $P_0(x)$ is selfadjoint by construction therefore this condition is fulfilled.
\item{\it (C3T):} $Q_0^\dag(x)S_2(x)$ should be selfadjoint. This condition implies to apply the relations for selfadjointness (see Appendix II) to such second order differential operator that implies:
 \begin{equation}
a^{(0,1)}(u_0,v_0)f^{(0,1,0)}(\underline{v})=\frac{d}{dx}\left[a^{(0,1)}(u_0,v_0)f^{(0,0,1)}(\underline{v})\right]\label{C3T_t}
\end{equation}
\item{\it (EM):} This condition is written from eq. (\ref{casoa}) using the expressions for $R_1$ and $R_2$ given by equations (\ref{R1R2dif}). It is a long expression that we do not write explicitely but it is of the form of a polynomial in $v_6, v_5, v_4, v_3$:
\begin{equation}
c_0(u_0,\underline{v})v_6+c_1(u_0,\underline{v})v_5+c_2(u_0,\underline{v})v_4+c_3(u_0,\underline{v})v^2_3+\ldots=0
\end{equation}
\end{itemize}
where $u_0=f(\underline{v})$. Both conditions are assumed to be identities that (as we explained above) are fulfill for any value of $v_n$ for $n\leq 2$. Therefore the polynomial structure on high derivatives imply that their coefficients should be zero. That gives us conditions on the functional forms of our unknowns: $a$ and $f$. It is convenient to be solving such conditions in an optimal way. We followed this line of reasoning:
\begin{itemize}
\item (1) Coefficient of $v_6$ from (EM) equal to zero:
\begin{equation}
a^{(1,0)}(u_0,v_0)f^{(0,0,1)}(\underline{v})^2\left(2 D(u_0)
-\chi(u_0)a^{(1,0)}(u_0,v_0)\right)=0
\end{equation}
We know that $a^{(1,0)}(u_0,v_0)\neq 0$ because the contrary would imply that $a$ only depends on $v_0$ that is against our initial assumption on $a$. Similarly $f^{(0,0,1)}(\underline{v})\neq 0$ by construction. Therefore from the last factor we find that
\begin{equation}
a(u_0,v_0)=\int du_0\frac{2D(u_0)}{\chi(u_0)}+\tilde a(v_0)\label{item1}
\end{equation}
We include this relation into conditions (CT3) and (EM) and we go to the next higher non-zero order.

\item (2) Coefficient of $v_3$ from (C3T) equal to zero: 
\begin{equation}
\tilde a^{(1)}(v_0)f^{(0,0,2)}(\underline{v})=0
\end{equation} 
 $\tilde a^{(1)}(v_0)f^{(0,0,2)}$ by construction and therefore $f$ should be a linear function of $v_2$:
\begin{equation}
f(\underline{v})=f_0(v_0,v_1)+v_2f_1(v_0,v_1)\label{y0}
\end{equation}
By using this relation, we rewrite the full (C3T) condition as:
\begin{equation}
v_1\frac{\partial}{\partial v_0}( \tilde a'(v_0)f_1(v_0,v_1))=\tilde a'(v_0)\frac{\partial f_0(v_0,v_1)}{\partial v_1}\label{C3T_fin}
\end{equation}
We use these relations into the condition (EM) and move on to the next non-zero order.
\item (3) Coefficient of $v_4$ from  (EM) equal to zero:
\begin{equation}
-4D(u_0)-2 u_0 D'(u_0)-g_0(v_0,v_1)D'(u_0)+g_1(v_0,v_1)\chi''(v_0)=0\label{y1}
\end{equation}
where
\begin{eqnarray}
g_0(v_0,v_1)&=&2\tilde a'(v_0)^{-1}\left(v_1^2\tilde a''(v_0)f_1(v_0,v_1)-f_0(v_0,v_1)\tilde a'(v_0)\right)\label{yc1}\\
g_1(v_0,v_1)&=&v_1f_1(v_0,v_1)\left(2E+v_1\tilde a'(v_0)\right)\label{yc2}
\end{eqnarray}
$u_0$ is arbitrary and independent on $v_0$ and $v_1$, therefore eq.(\ref{y1}) have three possible scenarios:
\begin{eqnarray}
&&\text{Case (a) } D(u_0)=D_0\quad,\quad g_1(v_0,v_1)=\Lambda_1=cte\nonumber\\
&&\text{Case (b) } \chi''(u_0)=0\quad,\quad g_0(v_0,v_1)=\Lambda_0=cte\nonumber\\
&&\text{Case (c) } g_0(v_0,v_1)=\Lambda_0=cte\quad,\quad g_1(v_0,v_1)=\Lambda_1=cte
\end{eqnarray}
Observe that we have not considered the possibility that  $D'(u_0)=c_0\chi''(u_0)$ because it is studied  in the $(n,m,l,s)=(0,0,0,0)$  case. At this point we should study each case separately:
\begin{itemize}
\item {\it Case (a):} $D(u_0)=D_0$, $g_1(v_0,v_1)=\Lambda_1=cte$  

Equation (\ref{y1}) implies:
\begin{equation}
\chi''(u_0)=\frac{4D_0}{\Lambda_1}\label{g1c}
\end{equation}
and therefore this case applies only when $D(u_0)=D_0$, $\chi(u_0)=k_0+k_1 u_0+k_2 u_0^2$ with $\Lambda_1=2D_0/k_2$. Equation (\ref{yc2}) implies :
\begin{equation}
f_1(v_0,v_1)=\frac{2D_0}{k_2v_1}\frac{1}{2E+v_1\tilde a'(v_0)}
\end{equation}
We use the remaining of condition (CT3), eq.(\ref{C3T_fin}), to get the form of $f_0(v_0,v_1)$:
\begin{equation}
f_0(v_0,v_1)=f_{00}(v_0)-\frac{4D_0E\tilde a''(v_0)}{k_2 \tilde a'(v_0)^2}\frac{1}{2E+v_1\tilde a'(v_0)}
\end{equation}
where $f_{00}(v_0)$ is an arbitrary function to be determined.
\item {\it Case (b):} $\chi''(u_0)=0$, $g_0(v_0,v_1)=\Lambda_0=cte$

 In this case $\chi(u_0)=k_0+k_1 u_0$ and equation (\ref{y1}) implies:
\begin{equation}
\frac{D'(u_0)}{D(u_0)}=\frac{-4}{\Lambda_0+2u_0}\Rightarrow D(u_0)=\frac{D_0}{(\Lambda_0+2u_0)^2}
\end{equation}
and
\begin{equation}
g_0(v_0,v_1)=\Lambda_0\Rightarrow f_0(v_0,v_1)=-\frac{\Lambda_0}{2}+\tilde a'(v_0)^{-1}v_1^2\tilde a''(v_0)f_1(v_0,v_1)
\end{equation}
We again use the remaining of the condition (C3T), eq.(\ref{C3T_fin}), to find $f_1(v_0,v_1)$:
\begin{equation}
f_1(v_0,v_1)=\tilde a'(v_0) f_{11}(v_1 \tilde a'(v_0))
\end{equation}
where $f_{11}(\beta)$ is and arbitrary function to be determined.
\item {\it Case (c):} $g_0(v_0,v_1)=\Lambda_0=cte$, $g_1(v_0,v_1)=\Lambda_1=cte$

From the definitions of $g_0$ and $g_1$ in eqs. (\ref{yc1}) and (\ref{yc2}) respectively,  we get the conditions:
\begin{eqnarray}
g_0(v_0,v_1)&=&\Lambda_0\Rightarrow f_0(v_0,v_1)=-\frac{\Lambda_0}{2}+\tilde a'(v_0)^{-1}v_1^2\tilde a''(v_0)f_1(v_0,v_1)\nonumber\\
g_1(v_0,v_1)&=&\Lambda_1\Rightarrow f_1(v_0,v_1)=\frac{\Lambda_1}{v_1}\frac{1}{2E+v_1\tilde a'(v_0)}\nonumber\\
-4D(u_0)&-&2u_0D'(u_0)-\Lambda_0 D'(u_0)+\Lambda_1\chi''(u_0)=0\nonumber
\end{eqnarray}
One can easily check that the condition (\ref{C3T_fin}) is fulfilled by these $f_0(v_0,v_1)$ and $f_1(v_0,v_1)$ functions.
\end{itemize}

\item We assume now that $\phi^*(x)=\tilde\phi^*(x)$. This choice is convenient because implies that the boundary conditions for the transformed field $\tilde\phi(x)$ are the same as the ones for $\phi(x)$ and therefore the second property in (\ref{coco1}) is immediately fulfilled. Then:
\begin{equation}
a(\phi^*(x),\phi^*(x))=0\Rightarrow \tilde a(\phi^*(x))=\bar a(\phi^*(x))\quad,\quad-\int^{w}dv\,\frac{2D(v)}{\chi(v)}
\end{equation}
where we have used (\ref{item1}). There are two possibilities: (a) $\tilde a(v_0)=\bar a(\phi^*(x))$ $\forall v_0$ or (b) $\tilde a(v_0)=\bar a(v_0)$ $\forall v_0$. The first case contradicts our $a$'s initial choice where we assumed that there weren't any explcit x-dependence on it. Therefore
\begin{equation}
a(u_0,v_0)=2\int_{v_0}^{u_0}du \frac{D(u)}{\chi(u)}\label{aeq}
\end{equation}

\item Another piece of information  is obtained from the differential equation that define the stationary state, that for this Diffusive System is given by:
\begin{equation}
\frac{d}{dx}\left[-D(\phi^*(x))\frac{d\phi^*(x)}{dx}+E\chi(\phi^*(x))\right]=0
\end{equation}
with the corresponding boundary conditions. We also know from eq.(\ref{y0}) that: 
\begin{equation}
\phi^*(x)=f_0(\phi^*(x),\phi^{*'}(x))+\phi^{*''}(x)f_1(\phi^*(x),\phi^{*'}(x))
\end{equation}
Obviously both differential equations should have the same solutions given the boundary conditions. Just by eliminating $\phi^{*''}(x)$ from both equations we get a relation between $f_0$ and $f_1$ at the stationary state that help us to determine the missing parts of $f$ in all three cases: 
\begin{itemize}
\item {\it Case (a):}  $f_{00}(v_0)=v_0$ and therefore:
\begin{equation}
f(v_0,v_1,v_2)=v_0-\frac{\chi(v_0)}{k_2 v_1}\frac{E\chi'(v_0)v_1-D_0v_2}{E\chi(v_0)-D_0v_1}
\end{equation}
\item {\it Case (b):}  In this case $\chi(v_0)=k_0+k_1 v_0$ that makes that the stationary state has the nice property that $D(\phi^*(x))\phi^{*'}(x)=E k_1\phi^*(x)+E k_0 -J$ where $J$ is the stationary current. Therefore  the unknown function $f_{11}(\beta)$ is given by:
\begin{equation}
f_{11}(\beta)=\frac{2\bar D_0}{\beta(2J-(\beta+2E)(k_0-k_1\bar\Lambda_0))}
\end{equation}
where $\bar D_0=D_0/4$ and $\bar\Lambda_0=\Lambda_0/2$ so $D(u)=\bar D_0/(\bar\Lambda_0-u)^2$. Finally we get:
\begin{eqnarray}
&&f(v_0,v_1,v_2)=\nonumber\\
&&-\bar\Lambda_0+\bar D_0\frac{(D(v_0)\chi'(v_0)-D'(v_0)\chi(v_0))v_1^2-D(v_0)\chi(v_0)v_2}{v_1 D(v_0)((k_0 E-J-k_1 E \bar\Lambda_0)\chi(v_0)-(k_0-k_1\bar\Lambda_0)D(v_0)v_1)}
\end{eqnarray}
\item {\it Case (c):} This order only is accomplished for situations already studied in (a) or (b). 
\end{itemize} 

\item The remaining condition (EM) is fulfilled once we use the full expressions for $a$ and $f$ for each nontrivial case. 
\end{itemize} 
We see that in this case  we have determined the precise form of $a$ and $f$ functions such that the conditions (C1T), (C3T) and (EM) are fulfilled for some concrete forms of $D(u)$ and $\chi(u)$. We can now to compute the quasipotential for each case.

\subsection{\bf (a) \boldmath$D(u)=D_0$, \boldmath$\chi(u)=k_0+k_1 u+k_2 u^2$}
We have found:
\begin{eqnarray}
A[\phi,\tilde\phi;x)&=&2D_0\int_{\tilde\phi(x)}^{\phi(x)}\frac{du}{\chi(u)}\equiv a(\phi(x),\tilde\phi(x))\nonumber\\
\phi[\tilde\phi;x]&=&\tilde\phi(x)-\frac{\chi(\tilde\phi(x))}{k_2\tilde\phi'(x)}\frac{E\chi'(\tilde\phi(x))\tilde\phi'(x)-D_0\tilde\phi''(x)}{E\chi(\tilde\phi(x))-D_0\tilde\phi'(x)}\label{rel70}
\end{eqnarray}
We use eq.(\ref{qp4}) to compute the quasipotential that it can be written as:
\begin{eqnarray}
V_0[\eta]&=&V_0[\phi^*]+I_1-I_2\nonumber\\
I_1&=&\int_0^1d\lambda\int_\Lambda dx\,(\tilde\eta(x)-\phi^*(x))\frac{\delta}{\delta\tilde\phi(x;\lambda)}\left[\int_\Lambda dy\,\int du_0\, a(u_0,v_0)\biggr\vert_{\substack{u_0=\phi[\tilde\phi(\lambda);y] \\v_0=\tilde\phi(y;\lambda)}}\right]\nonumber\\
I_2&=&\int_0^1d\lambda\int_\Lambda dx\,(\tilde\eta(x)-\phi^*(x))\int du_0\,\frac{\partial a(u_0,v_0)}{\partial v_0}\biggr\vert_{\substack{u_0=\phi[\tilde\phi(\lambda);x] \\v_0=\tilde\phi(x;\lambda)}}\label{int1}
\end{eqnarray}
where $\tilde\phi(x;\lambda)=\phi^*(x)+\lambda(\tilde\eta(x)-\phi^*(x))$,
and from eq.(\ref{rel70}) $\tilde\eta(x)$ is solution of the differential equation:
{\boldmath
\begin{equation}
\eta(x)=\tilde\eta(x)-\frac{\chi(\tilde\eta(x))}{k_2\tilde\eta'(x)}\frac{E\chi'(\tilde\eta(x))\tilde\eta'(x)-D_0\tilde\eta''(x)}{E\chi(\tilde\eta(x))-D_0\tilde\eta'(x)}\label{rel80}
\end{equation}}
for a given $\eta(x)$ field with given boundary conditions $\tilde\phi(x)=\phi(x)$ $\forall x\in\partial\Lambda$.

 and we have used the relation:
\begin{eqnarray}
\int_\Lambda dy\, a(\phi[\tilde\phi;y],\tilde\phi(y))\frac{\delta\phi[\tilde\phi;y]}{\delta\tilde\phi(x)}&=&\frac{\delta}{\delta\tilde\phi(x)}\int_\Lambda dy\int du_0\,a(u_0,v_0)\biggr\vert_{\substack{u_0=\phi[\tilde\phi;y] \\v_0=\tilde\phi(y)}}\nonumber\\
&-&\int du_0\,\frac{\partial a(u_0,v_0)}{\partial v_0}\biggr\vert_{\substack{u_0=\phi[\tilde\phi;y] \\v_0=\tilde\phi(y)}}
\end{eqnarray}

The first integral in (\ref{int1}) is just:
\begin{eqnarray}
I_1&=&\int_0^1d\lambda\frac{d}{d\lambda}\left[\int_\Lambda dy\,\int du_0\, a(u_0,v_0)\right]_{\substack{u_0=\phi[\tilde\phi(\lambda);y] \\v_0=\tilde\phi(y;\lambda)}}\nonumber\\
&=&\int_\Lambda dy\left[v(\eta(y))-\eta(y)v'(\eta(y))-v(\phi^*(y))+\phi^*(y)v'(\phi^*(y))\right]\label{I1}
\end{eqnarray}
where $v(u_0)=2D_0\int du_0\int du_0/\chi(u_0)$. 

The second integral needs a little more work. First, it can be written:
\begin{equation}
I_2=-2D_0\int_0^1d\lambda\int_\Lambda dx\,(\tilde\eta(x)-\phi^*(x))\frac{\phi[\tilde\phi(\lambda);x]}{\chi(\tilde\phi(x;\lambda))}
\end{equation}
This integral can be separated into two pieces: one without derivatives of $\tilde\phi(x;\lambda)$ and the other with its derivatives:
\begin{eqnarray}
I_2&=&I_{21}+I_{22}\nonumber\\
I_{21}&=&\int_{0}^1d\lambda\int_\Lambda dx\,\frac{\tilde\eta(x)-\phi^*(x)}{\chi(\tilde\phi(x;\lambda))}\left[\tilde\phi(x;\lambda)-\frac{1}{k_2}\chi'(\tilde\phi(x;\lambda))\right]\nonumber\\
I_{22}&=&\frac{D_0}{k_2}\int_{0}^1d\lambda\int_\Lambda dx\,(\tilde\eta(x)-\phi^*(x))\frac{1}{u(x;\lambda)\chi(\tilde\phi(x;\lambda))}\frac{u'(x;\lambda)}{E-D_0 u(x;\lambda)}
\end{eqnarray}
where $u(x;\lambda)=\tilde\phi'(x;\lambda)/\chi(\tilde\phi(x;\lambda))$. $I_{21}$ can be evaluated by making the change of variables $\lambda\rightarrow \tilde\phi(x;\lambda)$ at each $x$ and we get:
\begin{eqnarray}
I_{21}&=&\frac{1}{2D_0}\int_\Lambda dx\biggl[\tilde\eta(x)v'(\tilde\eta(x))-v(\tilde\eta(x))-\frac{2D_0}{k_2}\log\chi(\tilde\eta(x))\nonumber\\
&-&\phi^*(x)v'(\phi^*(x))+\frac{2D_0}{k_2}\log\chi(\phi^*(x))
\biggr]
\end{eqnarray}
 Integral $I_{22}$ can be written:
\begin{equation}
I_{22}=\frac{D_0}{k_2E}\int_{0}^1d\lambda\int_\Lambda dx\,\frac{\tilde\eta(x)-\phi^*(x)}{\chi(\tilde\phi(x;\lambda))}\frac{d}{dx}\log\left[\frac{\vert D_0 u(x;\lambda)\vert}{\vert E-D_0 u(x;\lambda)\vert}\right]
\end{equation}
and after integrating by parts on $x$:
\begin{equation}
I_{22}=-\frac{D_0}{k_2E}\int_{0}^1d\lambda\int_\Lambda dx\,\frac{d}{dx}\frac{\tilde\eta(x)-\phi^*(x)}{\chi(\tilde\phi(x;\lambda))}\log\left[\frac{\vert D_0 u(x;\lambda)\vert}{\vert E-D_0 u(x;\lambda)\vert}\right]
\end{equation}
we use now the relation: 
\begin{equation}
\frac{du(x;\lambda)}{d\lambda}=\frac{d}{dx}\frac{\tilde\eta(x)-\phi^*(x)}{\chi(\tilde\phi(x;\lambda))}
\end{equation}
to get
\begin{eqnarray}
I_{22}&=&-\frac{D_0}{k_2E}\int_\Lambda dx\,\int_{0}^1d\lambda\frac{du(x;\lambda)}{d\lambda}\log\left[\frac{\vert D_0 u(x;\lambda)\vert}{\vert E-D_0 u(x;\lambda)\vert}\right]\nonumber\\
&=&-\frac{D_0}{k_2E}\int_\Lambda dx\,\int_{u(x;0)}^{u(x;1)}du\log\left[\frac{\vert D_0 u\vert}{\vert E-D_0 u\vert}\right]
\end{eqnarray}
where $u(x,0)={\phi^*}'(x)/\chi(\phi^*(x))$ and $u(x,1)=\eta'(x)/\chi(\eta(x))$. After doing the integral of the logarithm and putting together all the pieces $I_1$, $I_{21}$ and $I_{22}$ we get the final expression for $V_0$:
{\boldmath
\begin{eqnarray}
V_0[\eta]&&=V_0[\phi^*]+\int_\Lambda dx\,\biggl[v(\eta(x))-v(\tilde\eta(x))\nonumber\\
&&-(\eta(x)-\tilde\eta(x))v'(\tilde\eta(x))
-\frac{2D_0}{k_2}\log\frac{\chi(\tilde\eta(x))}{\chi(\phi^*(x))}\nonumber\\
&&+\frac{2D_0}{k_2E}\biggl(\left(\frac{D_0\tilde\eta'(x)}{\chi(\tilde\eta(x))}-E\right)\log\biggl\vert\frac{D_0\tilde\eta'(x)}{\chi(\tilde\eta(x))}-E\biggr\vert-\frac{D_0\tilde\eta'(x)}{\chi(\tilde\eta(x))}\log\biggl\vert\frac{D_0\tilde\eta'(x)}{\chi(\tilde\eta(x))}\biggr\vert\nonumber\\
&&-\left(\frac{D_0{\phi^*}'(x)}{\chi(\phi^*(x))}-E\right)\log\biggl\vert\frac{D_0{\phi^*}'(x)}{\chi(\phi^*(x))}-E\biggr\vert+\frac{D_0{\phi^*}'(x)}{\chi(\phi^*(x))}\log\biggl\vert\frac{D_0{\phi^*}'(x)}{\chi(\phi^*(x))}\biggr\vert
\biggr)\biggr]\label{quasi3i}
\end{eqnarray}}
with $\tilde\eta(x)$ solution of the differential equation (\ref{rel80}) with boundary conditions $\tilde\eta(x)=\eta(x)$ $\forall x\in\partial\Lambda$.

We observe that the case $k_2\rightarrow 0$ seems to be singular; however, it is not. In order to do the limit let us think that $k_2$ is a perturbative parameter and then we assume that exist a well defined expansion: $\tilde\eta(x)=\tilde\eta_0(x)+k_2\tilde\eta_1(x)+\ldots$. We apply this expansion to eq. (\ref{rel80}), and it appears the order $k_2^{-1}$. Its coefficient should be zero and therefore:
\begin{equation}
-D_0\tilde\eta_0''(x)+Ek_1\tilde\eta_0'(x)=0\Rightarrow \tilde\eta_0(x)=\phi^*(x)
\end{equation}
where $\phi^*(x)$ is now the stationary state when $k_2=0$. For instance, $\phi^*(x)$ for the fixed boundary condition case 
($\phi^*(0)=\phi_0$ and $\phi^*(1)=\phi_1$)  is
\begin{equation}
\phi^*(x)=\frac{J}{k_1 E}+\frac{\phi_1-\phi_0}{e^{\tilde E}-1}e^{\tilde E x}\quad,\quad J=Ek_1 \frac{\phi_0 e^{\tilde E}-\phi_1}{e^{\tilde E}-1}
\end{equation}
where $\tilde E=Ek_1/D_0$ and $J$ is the current.
The order $k_2^0$ has the form:
{\boldmath
\begin{eqnarray}
\eta(x)&=&-\phi^*(x)\left(1+\frac{2D_0{\phi^*}'(x)}{k_0E+J}\right)\nonumber\\
&-&\frac{\chi_0(\phi^*(x))}{(k_0 E+J){\phi^*}'(x)}\left(k_1 E{\tilde\eta_1}'(x)-D_0{\tilde\eta_1}''(x)\right)\label{eer}
\end{eqnarray}}
and $\chi_0(u)=k_0+k_1 u$. This differential equation for $\tilde\eta_1(x)$ should be solved with boundary conditions $\tilde\eta_1(x)=0$ $\forall x\in\partial\Lambda$ because in the expansion $\tilde\eta_0$ carries the original boundary conditions. Finally, with all this information we can expand the quasipotential around $k_2=0$ and we get for
{\boldmath$D(u)=D_0$ and \boldmath$\chi(u)=k_0+k_1 u$}:
{\boldmath
\begin{eqnarray}
&&V_0[\eta]=V_0[\phi^*]+\int_\Lambda dx\,\biggl[v(\eta(x))-v(\phi^*(x))-(\eta(x)-\phi^*(x))v'(\phi^*(x))\nonumber\\
&&
+\frac{2D_0^2E}{\chi_0(\phi^*(x))^2}\left(\chi_0(\phi^*(x)){\tilde\eta_1}'(x)-k_1{\phi^*}'(x)\tilde\eta_1(x)\right)\log\left[\left\vert 1-\frac{E\chi_0(\phi^*(x))}{D_0{\phi^*}'(x)}\right\vert\right]\nonumber\\
&&-\frac{2D_0k_1\tilde\eta_1(x)}{\chi_0(\phi^*(x))}
\biggr]
\end{eqnarray}}
with $\tilde\eta_1(x)$ solution of eq.(\ref{eer}).
Observe that $\tilde\phi_1^*(x)\neq\phi^*(x)$. We can compute $\tilde\phi_1^*(x)$ from eq.(\ref{eer}) for the fixed boundary condition case:
\begin{eqnarray}
\tilde\phi_1^*(x)&=&\frac{\phi_1-\phi_0}{k_1}\frac{1}{(e^{\tilde E}-1)^2}\biggl[(\phi_1-\phi_0)\left(e^{\tilde E x}-1\right)\left(e^{\tilde E}-e^{\tilde E x}\right)\nonumber\\
&+&2\tilde E (\phi_0 e^{\tilde E}-\phi_1)\left[e^{\tilde Ex}(1-x)+\frac{e^{\tilde E x}-e^{\tilde E}}{e^{\tilde E}-1}\right]
\biggr]
\end{eqnarray}
that in the limit $\tilde E\rightarrow 0$ is reduced to
\begin{equation}
\tilde\phi_1^*(x)=\frac{2}{k_1}(\phi_1-\phi_0)^2x(1-x)
\end{equation}

\subsection{\bf (b) \boldmath$D(u)=\bar D_0/(\bar\Lambda_0+u)^2$, \boldmath$\chi(u)=k_0+k_1 u$}
We found in this case:
\begin{eqnarray}
A[\phi,\tilde\phi;x)&=&2\int_{\tilde\phi(x)}^{\phi(x)}\frac{du D(u)}{\chi(u)}\equiv a(\phi(x),\tilde\phi(x))\nonumber\\
\phi[\tilde\phi;x]&=&-\bar\Lambda_0\nonumber\\
&+&\frac{\bar D_0\left[(k_1D(\tilde\phi(x))-D'(\tilde\phi(x))\chi(\tilde\phi(x)))\tilde\phi'(x)^2-D(\tilde\phi(x))\chi(\tilde\phi(x))\tilde\phi''(x)\right]}{\tilde\phi'(x)D(\tilde\phi(x))\left[a_0\chi(\tilde\phi(x))-a_1 D(\tilde\phi(x))\tilde\phi'(x)\right]}
\end{eqnarray}
where $a_0=k_0 E-J-k_1 E \bar\Lambda_0$ and $a_1=k_0-k_1\bar\Lambda_0$.
We use these expressions in Eq. (\ref{qp4}), and initially, we follow similar steps as in the above case to get:
\begin{eqnarray}
&&V_0[\eta]=V_0[\phi^*]+\int_\Lambda dx\,\left[v(\eta(x))-v(\phi^*)-\eta(x) v'(\tilde\eta(x))+\phi^*(x) v'(\phi^*(x))\right]\nonumber\\
&&+\int_0^1d\lambda\,\int_\Lambda dx (\tilde\eta(x)-\phi^*(x))v''(\tilde\phi(x;\lambda))\phi[\tilde\phi;x]
\end{eqnarray}
where we remind that $v(u)=2\int du\int duD(u)/\chi(u)$. After substituting $\phi[\tilde\phi;y]$ from its expression we can decompose the last integral into two pieces:
\begin{equation}
\int_0^1d\lambda\,\int_\Lambda dx (\tilde\eta(x)-\phi^*(x))v''(\tilde\phi(x;\lambda))\phi[\tilde\phi;x]=-\bar\Lambda_0 I_{11}-\bar D_0 I_{12}
\end{equation}
where
\begin{eqnarray}
I_{11}&=&\int_0^1d\lambda\,\int_\Lambda dx (\tilde\eta(x)-\phi^*(x))v''(\tilde\phi(x;\lambda))\nonumber\\
I_{12}&=&\int_0^1d\lambda\,\int_\Lambda dx (\tilde\eta(x)-\phi^*(x))v''(\tilde\phi(x;\lambda))\frac{u'(x;\lambda)}{u(x;\lambda)(a_0+a_1 u(x;\lambda))}
\end{eqnarray}
where $u(x,\lambda)=v''(\tilde\phi(x;\lambda))\tilde\phi'(x;\lambda)/2$.
The integral $I_{11}$ is straightforward evaluated using the change of variables $\lambda\rightarrow\tilde\phi(x;\lambda)$ at each $x$. Then:
\begin{equation}
I_{11}=\int_\Lambda dx\,\left[v'(\tilde\eta(x))-v'(\phi^*(x))\right]
\end{equation}
The integral $I_{12}$ is done by first rewritting $u'/(u(a_0+a_1 u))=(\log\vert u/(a_0+a_1u)\vert)'/a_0$. Second, we integrate by parts where the surface term is zero due to the boundary conditions and third, we use the relation:
\begin{equation}
\frac{du(x;\lambda)}{d\lambda}=\frac{1}{2}\frac{d}{dx}\left[v''(\tilde\phi(x;\lambda)(\tilde\eta(x)-\phi^*(x))\right]
\end{equation}
Finally we put all the terms together and we find:
{\boldmath
\begin{eqnarray}
V_0[\eta]&=&V_0[\phi^*]+\int_\lambda dx\,\biggl[v(\eta(x))-v(\phi^*)-(\eta(x)+\bar\Lambda_0) v'(\tilde\eta(x))+(\phi^*(x)\nonumber\\
&+&\bar\Lambda_0) v'(\phi^*(x))
+\frac{\bar D_0}{a_0}v''(\phi^*(x)){\phi^*}'(x)\log\left[\left\vert a_1+\frac{2a_0}{v''(\phi^*(x)){\phi^*}'(x)}\right\vert\right]\nonumber\\
&-&\frac{\bar D_0}{a_0}v''(\tilde\eta(x)){\tilde\eta}'(x)\log\left[\left\vert a_1
+\frac{2a_0}{v''(\tilde\eta(x)){\tilde\eta}'(x)}\right\vert\right]\nonumber\\
&-&\frac{2\bar D_0}{a_1}\log\left[\left\vert\frac{2a_0+a_1v''(\tilde\eta(x))\tilde\eta'(x)}{2a_0+a_1v''(\phi^*(x)){\phi^*}'(x)} \right\vert\right]
\biggr]\label{quasi4i}
\end{eqnarray}}
 where $\tilde\eta(x)$ is solution of the differential equation:
 {\boldmath
 \begin{eqnarray}
 &&\eta(x)=-\bar\Lambda_0\nonumber\\
&&+\frac{\bar D_0\left[(k_1D(\tilde\eta(x))-D'(\tilde\eta(x))\chi(\tilde\eta(x)))\tilde\eta'(x)^2-D(\tilde\eta(x))\chi(\tilde\eta(x))\tilde\eta''(x)\right]}{\tilde\eta'(x)D(\tilde\eta(x))\left[a_0\chi(\tilde\eta(x))-a_1 D(\tilde\eta(x))\tilde\eta'(x)\right]}
 \end{eqnarray}}
 for any given $\eta(x)$ field.

\subsection*{\bf(iii) \boldmath$(n,m,l,s)=(0,0,1,1)$}

In this case we choose $a=a(\phi(x),\tilde\phi(x))$ and $\phi^{(1)}(x)=f(\phi(x),\tilde\phi(x),\tilde\phi'(x))$. The corresponding operators (eqs. (\ref{P},\ref{Q},\ref{L},\ref{S})) have the form:
\begin{eqnarray}
P_0(x)&=&a^{(1,0)}(u_0,v_0)\nonumber\\
Q_0(x)&=&a^{(0,1)}(u_0,v_0)\nonumber\\
L_1(x)&=&f^{(1,0,0)}(u_0,v_0,v_1)-\frac{d}{dx}\nonumber\\\nonumber\\
S_1(x)&=&f^{(0,1,0)}(u_0,v_0,v_1)+f^{(0,0,1)}(u_0,v_0,v_1)\frac{d}{dx}
\end{eqnarray} 
The conditions for the existence of $V_0$ are given by eq.(\ref{casob}):
\begin{itemize}
\item {\it (C1T):} This conditon is fulfilled because $P_0(x)$ is selfadjoint by construction.
\item {\it (C3T) and (EM):}  they are polynomials on $v_k$.
\end{itemize}
 The strategy we have followed to unveil the form of the funtions $a$ and $f$ is:
\begin{itemize}
\item (1) Coefficient of $v_2$ from (C3T) equal to zero:
\begin{equation}
f^{(0,0,2)}(u_0,v_0,v_1)=0 \Rightarrow f(u_0,v_0,v_1)=f_0(u_0,v_0)+v_1 f_1(u_0,v_0)\label{zz1}
\end{equation}
\item (2) Rest of (C3T) after substituying eq.(\ref{zz1}):
\begin{equation}
f_0(u_0,v_0)f_1^{(1,0)}(u_0,v_0)=f_0^{(0,1)}(u_0,v_0)+f_1(u_0,v_0)f_0^{(1,0)}(u_0,v_0)\label{C3Tff}
\end{equation}
\item (3) Coefficient $v_3$ from (EM): 

There are two possibilities, 
\begin{eqnarray}
f_1(u_0,v_0)&=&-\frac{a^{(0,1)}(u_0,v_0)}{a^{(1,0)}(u_0,v_0)}\nonumber\\
f_1(u_0,v_0)&=&\frac{\chi(u_0)a^{(0,1)}(u_0,v_0)}{2D(u_0)-\chi(u_0)a^{(1,0)}(u_0,v_0)}\nonumber
\end{eqnarray}
One can show that the first case implies the case $(n,m,l,s)=(0,0,0,0)$.
\item (4) Coefficient $v_2$ from (EM): 

We get a partial differential equation on $f_0(u_0,v_0)$ that, together eq.(\ref{C3Tff}) implies:
\begin{eqnarray}
a(u_0,v_0)&=&\tilde a(u_0)-\tilde a(v_0)\nonumber\\
f_0(u_0,v_0)&=&c_0\frac{\chi(u_0)}{2D(u_0)-\chi(u_0)\tilde a'(u_0)}\nonumber\\
f_1(u_0,v_0)&=&-\frac{\tilde a'(v_0)\chi(u_0)}{2D(u_0)-\chi(u_0)\tilde a'(u_0)}
\end{eqnarray} 
These relations implies that (C3T) is fulfilled.
\item (5) Rest of condition (EM):
\begin{equation}
c_0(c_0-2E)\chi(u_0)^2(c_0-v1\tilde a'(v_0))\chi''(u_0)=0\label{last}
\end{equation}
There are three possibilities: $c_0=0$, $c_0=2E$ or $\chi''(u_0)=0$. In order to elucidate which one of these is the correct, we apply our equation to the stationary state. First we know that  $\phi^*(x)=\tilde\phi^*(x)$ because $a(\phi^*(x),\tilde\phi^*(x))=0$.  Then, we apply this to our equation $u_1=f(u_0,v_0,v_1)$ and  we find that
$c_0=2\phi^{*'}(x)D(\phi^*(x))/\chi(\phi^*(x))$. By other hand we know that the stationary state is solution of $\phi^{*'}(x)D(\phi^*(x))/\chi(\phi^*(x))=E-J/\chi(\phi^*(x))$ that implies $\chi(u_0)=\chi_0$. This is coherent with one of the conditions in eq.(\ref{last}) and (EM) is fulfilled.  Finally, we can write eq.(\ref{zz1}) as:
\begin{equation}
\phi'(x)[2D(\phi(x))-\chi_0\tilde a'(\phi(x))]=2(E\chi_0-J)-\chi_0\tilde\phi'(x)\tilde a'(\tilde\phi(x))\label{218}
\end{equation}
that it can be integrated:
\begin{equation}
\tilde a(\phi(x))-\tilde a(\tilde\phi(x))=\frac{2}{\chi_0}\int_{\phi^*(x)}^{\phi(x)}du D(u)\label{219}
\end{equation}

The quasipotential can be calculated following similar steps as in the above cases. We know that $V_0$ can be written as:
\begin{equation}
V_0[\eta]=V_0[\phi^*]+I_1-I_2
\end{equation}
where
\begin{eqnarray}
I_1&=&\int_\Lambda dx\,\left[\int_{\phi^*(x)}^{\eta(x)}du\,\tilde a(u)-\eta(x)\tilde a(\tilde\eta(x))+\phi^*(x)\tilde a(\phi^*(x))\right]\nonumber\\
I_2&=&\int_0^1d\lambda\,\int_\Lambda dx\,\left(\eta(x)-\phi^*(x)\right)\phi(x;\lambda)\tilde a'(\tilde\phi(x;\lambda))
\end{eqnarray}
$I_2$ can be simplified by using eq.(\ref{219}) with $\tilde\phi(x;\lambda)$ and $\phi(x;\lambda)$ and after doing a $\lambda$ derivative in it we get the relation:
\begin{equation}
\tilde a'(\tilde\phi(x;\lambda))\left(\eta(x)-\phi^*(x)\right)=\frac{d\phi(x;\lambda)}{d\lambda}\left(\tilde a'(\phi(x;\lambda))-\frac{2}{\chi_0}D(\phi(x;\lambda))\right)
\end{equation}
Finally, we do the change of variables $\lambda\rightarrow\phi(x;\lambda)$ at each $x$ and we find:
\begin{equation}
I_2=\int_\Lambda dx\,\int_{\phi^*(x)}^{\eta(x)}du\,u\left[\tilde a'(u)-\frac{2}{\chi_0}D(u) \right]
\end{equation}
Putting together $I_1$ and $I_2$ we get:
{\boldmath
\begin{equation}
V_0[\eta]=V_0[\eta^*]+\frac{2}{\chi_0}\int_\Lambda dx\,\int_{\phi^*(x)}^{\eta(x)}du\,\int_{\phi^*(x)}^u dv D(v)\label{quasi5i}
\end{equation}}
for any $D(u)$ and {\boldmath $\chi(u)=\chi_0$}. $\phi^*(x)$ is solution of 
\begin{equation}
D(\phi^*(x)){\phi^*}'(x)=\chi_0 E-J
\end{equation}
\end{itemize}

Let us finish this section by commenting that we also attempted other values for $(n,m,l,s)$. For instance, $(0,1,0,1)$ reduces to case $(0,0,1,1)$ we have studied explicitly. However, we couldn't find anything for cases $(0,0,2,2)$, $(0,0,0,4)$, and $(0,2,0,2)$.

\section{VI. Quasipotentials for some one dimensional reaction-diffusion models}

We first study the reaction-diffusion model whose  Langevin equation is given by (\ref{rd}) with
\begin{equation}
F[\phi;x]=g(\phi)\phi''(x)+w(\phi(x))
\end{equation}
We choose {\boldmath $(n,m,l,s)=(0,0,0,0)$}. As it happened in the similar case of Diffusion Dynamics, conditions (C1T) and (C3T) are fulfilled by construction. The condition (EM) is built by using the functionals:
\begin{eqnarray}
R_1[\phi,\tilde\phi;x]&=&g(\phi)\phi''(x)+w(\phi(x))+a(x) h^2(\phi(x))\nonumber\\
R_2[\phi,\tilde\phi;x]&=&-a(x)\left[g'(\phi(x))\phi''(x)+w'(\phi(x))+a(x)h(\phi(x))h'(\phi(x))\right]\nonumber\\
&-&(a(x)g(\phi(x)))''
\end{eqnarray}
where $a(x)=a(\phi(x),\tilde\phi(x))$. After substitutions and some trivial algebra,  the condition (EM) becomes a polynomial on $v_k$ where their coefficients are equaled to zero:
\begin{itemize}
\item Coefficient of $v_2$ from (EM) equal to zero:
\begin{equation}
a(u_0,v_0;u_s)=\frac{C}{g(u_0)}
\end{equation}
This is a singular case because at the stationary state $a=0$ and therefore $g(u_s)^{-1}=0$. Let us assume that 
\begin{equation}
g(u)=\bar g(u) (u-u_s)^{-\alpha}\quad \alpha>0
\end{equation}
This is coherent whenever the stationary state is a constant: $\phi^*(x)=\phi^*$ (because we are not considering in this paper local functions of $g$, $w$ or $h$). Observe that with this election the deterministic evolution equation is: 
\begin{equation}
\dot\phi_D(x,t)=\bar g(\phi_D(x,t))(\phi_D(x,t)-\phi^*)^{-\alpha}\phi_D''(x,t)+w(\phi_D(x,t))
\end{equation}
Near the equilibrium, $\phi(x,t)=\phi^*+\epsilon(x,t)$ the dominant terms of this equation for very small values of $\epsilon$ can be written:
\begin{equation}
\partial_t\epsilon(x,t)=w'(\phi^*)\epsilon(x,t)+\epsilon^{-\alpha}\epsilon(x,t)^{-\alpha}\bar g(\phi^*)\partial_{xx}^2\epsilon(x,t)
\end{equation}
where $\phi^*$ is solution of $w(\phi^*)=0$. The diffusion term is singular but it goes very fast to zero because it behaves as if the system had almost an infinite diffusivity and, therefore, it homogenizes any initial profile very fast in such a way that the spatial second derivative becomes zero. The reaction term makes $\epsilon$ to evolve exponentially fast towards zero  whenever $w'(\phi^*)<0$. Therefore this singular system is well behaved near the stationary solution.

\item The rest of the  (EM) condition is fulfilled when:
\begin{equation}
h^2(u)=-2 g(u)w(u)
\end{equation}
One can check that with the form of $a$ and $h$, the hamiltonian $H[\phi,\pi]=0$ along the trajectory.
\end{itemize}
The quasipotential can we written as
\begin{equation}
V_0[\eta]=V_0[\phi^*]+C\int_0^1d\lambda\,\int_\Lambda dx\,\frac{(\tilde\eta(x)-\phi^*(x))}{g(f(\tilde\phi(x;\lambda)))}\frac{\partial f(v;\phi^*)}{\partial v}\biggr\vert_{v=\tilde\phi(x;\lambda)}
\end{equation}
where $\phi[\tilde\phi;x]=f(\tilde\phi(x);\phi^*(x))$ and $\tilde\phi(x;\lambda)=\phi^*(x)+\lambda(\tilde\eta(x)-\phi^*(x))$. We do the change of variable $\lambda\rightarrow\tilde\phi(x;\lambda)$ at each $x$ point and we get
{\boldmath
\begin{equation}
V_0[\eta]=V_0[\phi^*]+\int_\Lambda dx\,\int_{\phi^*}^{\eta(x)}\frac{du}{g(u)}
\end{equation}}
where
{\boldmath$g(\phi)=\bar g(\phi)(\phi-\phi^*)^{-\alpha}$, \boldmath$\alpha>0$}.

The other interesting reaction-diffusion model we have studied is the Poissonian Reaction-Diffusion Dynamics. We already defined this model on Section II.  We choose {\boldmath $(n,m,l,s)=(0,0,0,0)$}. Again, conditions (CT1) and (CT3) are fulfilled by construction. And the (EM) condition is build with 
 \begin{eqnarray}
R_1[\phi,\tilde\phi;x]&=&\phi''(x)-2(a'(x)\phi(x)(1-\phi(x)))'+b(\phi(x))e^{a(x)}-d(\phi(x))e^{-a(x)}\nonumber\\
R_2[\phi,\tilde\phi;x]&=&-a''(x)-a'(x)^2(1-2\phi(x))+b'(\phi(x))(1-e^{a(x)})\nonumber\\
&+&d'(\phi(x))(1-e^{-a(x)})
\end{eqnarray}
The coefficient for $v_2$ equaled to zero give us the first condition: 
\begin{equation}
f'(v_0)-f(v_0)(1-f(v_0))\frac{da(f(v_0),v_0)}{dv_0}=0\Rightarrow a(u_0,v_0)=\log\frac{u_0}{1-u_0}+C
\end{equation}
where $C$ is a constant that is fixed by the condition: $a(\phi^*,\phi^*)=0$: $C=\log[(1-\phi^*)/\phi^*]$. Let us remind that we are using periodic boundary conditions and that the stationary state is homogeneous: $\phi^*(x)=\phi^*$, where $\phi^*$ is solution of $b(\phi^*)=d(\phi^*)$.

We use the result on $a$ to we get the rest of the (EM) condition:
\begin{eqnarray}
&&u_0^2(1-\phi^*)^2b(u_0)-{\phi^*}^2(1-u_0)^2d(u_0)\nonumber\\
&&=u_0(1-u_0)(u_0-\phi^*)\left[\phi^*(1-u_0)d'(u_0)-u_0(1-\phi^*)b'(u_0)\right]
\end{eqnarray}
The solution of this equation is:
\begin{equation}
b(u)=\phi^*(1-u)h(u)\quad ,\quad d(u)=(1-\phi^*)u h(u)
\end{equation}
where $h(u)$ is a positive function. 

Once we know $a(u_0,v_0)$ and the functions $b(u)$ and $d(u)$ that fulfills the conditions we can compute the quasipotential. We follow almost the same steps as we did for the Diffusion case for $(n,m,l,s)=(0,0,0,0)$ and we get
{\boldmath
\begin{equation}
V_0[\eta]=V_0[\phi^*]+\int_\Lambda dx\,\left[\eta(x)\log\left[\frac{\eta(x)}{\phi^*}\right]+(1-\eta(x))\log\left[\frac{1-\eta(x)}{1-\phi^*}\right]\right]
\end{equation}
}
 This case was already studied by Gabrielli et al (1997) \cite{gabrielli,Bertini}.
 
We also studied case $(n,m,l,s)=(0,0,0,2)$, but no situation fulfilled the conditions.

\section{Summary of Results and Conclusions}

The mesoscopic description of non-equilibrium systems given by the MFT \cite{Bertini} is a solid background to study their generic properties. Even though MFT is mathematically simpler than its microscopic original description, it is still difficult to extract precise information from our actual analytical tools. This paper intends to build a method to get the stationary measure represented by the quasipotential at the small noise limit.  Formally, the quasipotential is obtained by a time integral of some variables along a path defined by a Hamiltonian that depends on the studied system, $(\phi(t),\pi(t))$:
\begin{equation}
V_0[\eta]\sim\int_{-\infty}^0 dt\int_{x\in\Lambda}\pi(x,t)\partial_t\phi(x,t)\label{ti}
\end{equation}
After a canonical transformation of type 1 with generator $L[\phi,\tilde\phi]$, the situation does not change too much, and the quasipotential still is mainly a time integral like eq. (\ref{ti}) along a path defined by the same Hamiltonian expressed in the new variables $(\tilde\phi(t),\tilde\pi(t))$.
However, we can effectively deform this last path into a straight line by using the properties of the canonical transformation and assuming that there exists a map between the paths followed by the original fields and the transformed ones:
\begin{equation}
\text{\bf Main Assumption:}\quad \phi(x,t)=\phi[\tilde\phi(t);x]\quad\forall t\in[-\infty,0]
\end{equation}
Therefore, the quasipotential can be expressed in the {\it convenient form}, eq.(\ref{qp4}):
\begin{eqnarray}
 V_0[\eta]&=&V_0[\phi^*]\nonumber\\&+&\int_{0}^1d\lambda\int_{\Lambda}dx\,(\tilde\eta(x)-\tilde\phi^*(x))\int_{\Lambda}dy\,A[\phi[\tilde\phi(\lambda)],\tilde\phi(\lambda);y]K[\tilde\phi(\lambda);y,x]
  \label{qp4b}
 \end{eqnarray}
 where $\tilde\phi(x,\lambda)=\tilde\phi^*(x)+\lambda(\tilde\eta(x)-\tilde\phi^*(x))$.   In this way, we manage to get rid of highly nontrivial path integration. 
 However, we pay the price of introducing two local functionals: $A[\phi;\tilde\phi;x]=\delta L[\phi,\tilde\phi]/\delta\phi(x)$ and $K[\tilde\phi;x,y]=\delta\phi[\tilde\phi;x]/\delta\tilde\phi(y)$.
 At this point, our method's goal is to design a way to determine directly both functionals: $A[\phi;x]$ and $K[\tilde\phi;x,y]$. We know that they depend on the canonical transformation and on Hamilton's equations that define the paths. Therefore, they cannot take any functional form. In fact, we show that they should fulfill three compatibility conditions (C1T), (C3T), and (EM) given by Eqs. (\ref{C1T}),  (\ref{C3T}) and  (\ref{EM}) respectively. These conditions are the core of our method: we initially give functional forms for $A$ and $\phi[\tilde\phi]$, using the compatibility conditions to determine their fine structure. 
 
 In this paper (Section IV) we only use  the family of functionals given by eqs.(\ref{a}) and (\ref{f}):
 \begin{equation}
A[\phi,\tilde\phi;x]=a(\phi(x),\phi'(x),\ldots,\phi^{(n)}(x),\tilde\phi(x),\tilde\phi'(x),\ldots,\tilde\phi^{(m)}(x),\phi^*(x))\label{ab}
\end{equation}
\begin{equation}
\phi^{(l)}(x)=f(\phi(x),\phi'(x),\ldots,\phi^{(l-1)}(x),\tilde\phi(x),\tilde\phi'(x),\ldots,\tilde\phi^{(s)}(x),\phi^*(x))\label{fb}
\end{equation}
Observe that $K$-functional can be derived from eq.(\ref{fb}) just by a functional derivative on both sides (see main text). This family permits to write the compatibility conditions compactly by using differential operators (see Eqs. (\ref{C1T0},\ref{C3T0},\ref{EMv1})) where some inverse operators appear. This fact makes their algebraic use very difficult, if not impossible. To go forward, we restrict ourselves to cases where such inverse differential operators are just functions. Then the compatibility conditions simplify and become eq.(\ref{casoa}) or (\ref{casob}), and they are algebraically manageable. 
 
 We show in the paper (sections III.4, V and VI) how to algebraically deal with such conditions once we fix  a particular set of $(n,m,l,s)$ values for the functions $a$ and $f$ (Eqs.(\ref{ab}) and (\ref{fb}) respectively). We observe that the compatibility conditions can determine the form of the functions $a$ and $f$. At the same time, they select a subfamily of Langevin equations (for instance, the functional form for the diffusion or the mobility for diffusive systems).  We explicitly apply this method to already well-known cases, discovering new solutions that may be of general interest. In particular, for one dimensional Diffusive Systems that are characterized by the functions $D(\phi)$ and $\chi(\phi)$ (see main text) we get the explicit quasipotential for the cases: (i) $D(u)=c\chi'(u)$  in eq.(\ref{quasii}), (ii) $D(u)=D_0$, $\chi(u)=k_0+k_1 u+k_2 u^2$ in eq. (\ref{quasi3i}), (iii) $D(u)=D_0/(\Lambda_0+u)^2$, $\chi(u)=k_0+k_1 u$ in eq.(\ref{quasi4i}) and (iv) $D(u)$, $\chi(u)=\chi_0$ in eq.(\ref{quasi5i}). 
 
 We observe sets of values $(n,m,l,s)$ that do not fulfill the compatibility conditions. In general, this method does not guarantee a priori that any given functional structure for $a$ and $f$ should be associated with a well-defined canonical transformation. Therefore this is a kind of trial-error method at this moment. It would be relevant in the future to have an {\it a priori} deeper knowledge about compatible functional forms for $A$ and $\phi[\tilde\phi]$. 
 
We think that this method may be developed and improved further in several ways. For instance, applying it to higher dimensional systems, at least initially, for simple cases as the zero-range model or even the SSE could be interesting. On the other hand, in the paper, we focused on transformations whose compatibility conditions do not contain generic differential operators inverse. For these cases, we handle the structure of the compatibility conditions easily. We think that there is a vast work to be done dealing with the more generic cases where such inverse operators appear. Probably they carry stronger non-local properties necessary to describe the behavior of more complex situations. It could also be interesting to define some other functional structures for $A$ and $\phi[\tilde\phi;x]$ as integrals of functions over local domains or similar. 

In this paper, we only dealt with canonical transformations of type I, and the study of other types may imply new quasipotential structures. Finally, it could be interesting to set up a systematic perturbation theory similar to the Bouchet et al. \cite{Bouchet} but associated with the canonical transformation we have presented.

\section{Acknowledgements}
 We acknowledge financial support from the Spanish "Ministerio de Ciencia e Innovación” and the 
"Agencia Estatal de Investigación (AEI)” under Project Ref. PID2020-113681GB-I00 as well as 
  the Consejer{\'\i}a de Conocimiento, Investigaci{\'o}n Universidad, Junta de Andaluc{\'\i}a and European Regional Development Fund,  Ref. A-FQM-175-UGR18  and in part by AFOSR [grant FA-9550-16-1-0037].

\section*{Appendix I}

Let $R[\phi;x]$ be a given local functional on $\phi$. We know that 
\begin{equation}
\frac{\delta B[\phi]}{\delta\phi(x)}=R[\phi;x]\label{a1}
\end{equation}
and let us assume that $B[\phi]$ exists, that is
\begin{equation}
\frac{\delta R[\phi;x]}{\delta\phi(y)}=\frac{\delta R[\phi;y]}{\delta\phi(x)}\label{a3}
\end{equation}
Then we can show that
\begin{equation}
B[\phi]=B[\phi^{*}]+\int_{0}^1 d\lambda\,\int_\Lambda dx\,(\phi(x)-\phi^{*}(x))R[\phi(\lambda);x]\label{a2}
\end{equation}
where
\begin{equation}
\phi(x;\lambda)=\phi^{*}(x)+\lambda (\phi(x)-\phi^{*}(x))
\end{equation}
{\it Demonstration:} Let us show that the derivative of (\ref{a2}) is (\ref{a1}).
\begin{equation}
\frac{\delta B[\phi]}{\delta\phi(y)}=\int_{0}^1 d\lambda\,\int_\Lambda dx\,\left[R[\phi(\lambda);x]\delta(x-y)+(\phi(x)-\phi^{*}(x))\frac{\delta R[\phi(\lambda);x]}{\delta\phi(y)}\right]
\end{equation}
We can use the relations:
\begin{eqnarray}
\frac{\delta R[\phi(\lambda);x]}{\delta\phi(y)}&=&\lambda\frac{\delta R[\phi;x]}{\delta\phi(y)}\biggr\vert_{\phi=\phi(\lambda)}\nonumber\\
\frac{dR[\phi(\lambda);x]}{d\lambda}&=&\int_{\Lambda}dy\,\frac{\delta R[\phi;y]}{\delta\phi(x)}\biggr\vert_{\phi=\phi(\lambda)}(\phi(y)-\phi^{*}(y))
\end{eqnarray}
where we have made use of (\ref{a3}). Then we get
\begin{equation}
\frac{\delta B[\phi]}{\delta\phi(y)}=\int_{0}^1 d\lambda\,\left[R[\phi(\lambda);y]+\lambda\frac{dR[\phi(\lambda);y]}{d\lambda}\right]=R[\phi(\lambda);y]\quad\text{c.q.d.}
\end{equation}

\section*{Appendix II:  Self adjoint conditions for a n-differential operator in one dimension}

Let's $L$ be a linear differential operator of n-th order:
\begin{equation}
L=\sum_{k=0}^na_k(x)\frac{d^k}{dx^k}
\end{equation}
where $a_k(x)\in\mathcal{R}\quad \forall x\in\Lambda\subset\mathcal{R}$. We define the inner product for two real analytic functions $u(x)$, $v(x)$:
\begin{equation}
\langle u,v\rangle=\int_\Lambda dx\,u(x)v(x)\label{inn}
\end{equation}
The adjoint of $L$, $L^\dagger$, is then defined by
\begin{equation}
\langle L^\dagger u,v\rangle=\langle u,Lv\rangle
\end{equation}
Therefore
\begin{equation}
 L^\dagger =\sum_{k=0}^n(-1)^k\frac{d^k}{dx^k} a_k(x)
\end{equation}
where there are assumed that the set of real functions $v(x)$, $u(x)$, where $L$ and $L^\dagger$ respectively apply have  boundary conditions such that
\begin{equation}
\sum_{k=1}^n\sum_{l=0}^{k-1}(-1)^l\frac{d^l}{dx^l}(a_k(x)u(x))\frac{d^{k-1-l}}{dx^{k-1-l}}v(x)\biggr\vert_{\partial\Lambda}=0\label{boun}
\end{equation}
$L$ is called {\it self-adjoint} if $L=L^\dagger$ and the set of boundary conditions for the $v(x)$ and $u(x)$ functions coincide and fulfills eq.(\ref{boun}). Therefore, the coefficients of $L$ such that $L=L^\dagger$ should be related by
\begin{equation}
a_k(x)=\sum_{l=k}^n(-1)^l\binom{l}{k}\frac{d^{l-k}}{dx^{l-k}}a_l(x)\quad k=1,\ldots,n\label{rel7}
\end{equation}
Observe that only the operators $L$ with $n$ even can be self-adjoint. That can be shown by applying relation (\ref{rel7}) to the case $k=n$. 

One realizes that not all the $n$-relations defined by (\ref{rel7}) are independent. In fact, we can show that the independent set of relations that define a self-adjoint operator is given by:
\begin{equation}
a_{2l+1}(x)=\frac{1}{(2l+1)!}\sum_{s=0}^{m-l-1}(2(s+l+1))!\,c_s\frac{d^{2s+1}}{dx^{2s+1}}a_{2l+2s+2}(x)\label{rel8}
\end{equation}
where $l=0,\ldots,m-1$ and $n=2m$. 
The  $c$'s are a set of numbers generated by the recurrence:
\begin{equation}
c_l=\frac{1}{2(2l+1)!}-\frac{1}{2}\sum_{k=0}^{l-1}\frac{c_k}{(2l-2k)!}\quad,\quad l>0\quad,\quad c_0=1/2
\end{equation}
For instance: $c_1=-1/24$, $c_2=1/240$, $c_3=-17/40320$, $c_4=31/725760$, $...$. Curiously enough we find that $c$'s follow another relation:
\begin{equation}
\frac{1}{(2l+2)!}=\sum_{k=0}^l\frac{c_k}{(2l-2k+1)!}
\end{equation}
We have computed the first one hundred values of $c$'s and found that they alternate signs and their modulus decrease exponentialy fast: $\vert c_l\vert\simeq 1.82 \exp[-2.29 l]$.

As an example,for $n=2$  the condition for self-adjointness is:
\begin{equation}
a_1(x)=\frac{da_2(x)}{dx}
\end{equation}
and for $n=4$ we have two conditions:
\begin{equation}
a_1(x)=\frac{da_2(x)}{dx}-\frac{d^3a_4(x)}{dx^3}\quad,\quad a_3(x)=2\frac{da_4(x)}{dx}
\end{equation}

Let us define the Green functions $G$ and $G^\dagger$ solutions of the equations:
\begin{equation}
LG(x,x_0)=\delta(x-x_0)\quad ,\quad L^\dagger G^\dagger(x,x_0)=\delta(x-x_0)
\end{equation}

\underline{\it Proposition:} $L$ is self-adjoint if and only if $G(x_1,x_2)=G(x_2,x_1)$

To prove the proposition let us choose $u(x)=G^\dagger(x,x_2)$ and $v(x)=G(x,x_1)$ for the inner product in eq. (\ref{inn}). Then we get:
\begin{equation}
 G^\dagger(x_1,x_2)=G(x_2,x_1)\label{inn2}
\end{equation}
If $L$ is self-adjoint then $G^\dagger(x_1,x_2)=G(x_1,x_2)$ and using eq.(\ref{inn2}) we prove the right implication.
If we assume $G(x_1,x_2)=G(x_2,x_1)$  we see from (\ref{inn2}) that $G^\dagger(x_1,x_2)=G(x_1,x_2)$. Assuming that there is an unique solution for each operator we get $L^\dagger=L$.

\end{document}